\documentclass[a4paper,11pt,titlepage,oneside]{book} 

\newcommand{\totalusecases}{89~}

\newcommand{\openusecases}{32~}
\newcommand{\whiteusecases}{23~}
\newcommand{\greyusecases}{28~}
\newcommand{\scientificusecases}{6~}
\newcommand{\characteristics}{24~}

\usepackage[T1]{fontenc}
\usepackage[utf8]{inputenc}
\usepackage{ae}               
\usepackage[pdftex]{graphicx}
\usepackage{vmargin}          
\usepackage{fancyhdr}         
\usepackage{subfigure}
\usepackage[hyphens]{url}
\usepackage{xcolor}
\usepackage{footnote}
 
\usepackage[absolute,overlay]{textpos}
\usepackage{tikz}
\usepackage[english]{babel}
\usepackage{color}
\usepackage{longtable}
\usepackage[bottom]{footmisc}
\usepackage[section,subsection,subsubsection]{placeins}
\usepackage{multirow}
\usepackage{multicol}
\usepackage{array}
\newcolumntype{P}[1]{>{\raggedright\arraybackslash}p{#1}}
\usepackage{nomencl}
\usepackage{color}
\usepackage{setspace}
\usepackage{amsmath}
\usepackage{amssymb}

\makenomenclature

\makeglossary

\definecolor{darkred}{rgb}{0.5,0,0}
\definecolor{darkgreen}{rgb}{0,0.8,0}
\definecolor{darkblue}{rgb}{0,0,0.5}

\setmarginsrb
{2.5cm}   
{1.5cm}  
{2.5cm}  
{2cm}  
{20pt}    
{0.25in}  
{9pt}    
{0.3in}

\usepackage{chngcntr}
\counterwithin{figure}{section}
\counterwithin{table}{section}

\newcommand{\forget}[1]{}  
\newcommand{\sandbox}[1]{} 
\newcommand{\longer}[1]{} 

\newcommand{\shorten}[1]{}

\usepackage{authblk}

\author[1]{Tentative author list\footnote{SPEC Confidential. This document should be considered confidential unless labeled otherwise.}}

\usepackage{multirow}
\usepackage{array}
\usepackage{ragged2e}
\usepackage{hyperref}
\usepackage[numbers]{natbib}
\usepackage{appendix}

\usepackage{balance}

\usepackage[utf8]{inputenc}
\usepackage{amsfonts}
\usepackage{slashbox,booktabs,amsmath}
\usepackage{amssymb}
\usepackage{pifont}
\usepackage{cleveref}

\usepackage{enumitem}

\newlist{inlinelist}{enumerate*}{1}
\setlist*[inlinelist,1]{
  label=(\roman*),
}

\usepackage{array, makecell}

\usepackage{tikz}
\usetikzlibrary{intersections}
\usepackage[normalem]{ulem}
\usepackage{hyperref}
\usepackage{mathtools}
\usepackage{mdframed}
\usepackage{tcolorbox}
\tcbuselibrary{skins}

\usepackage{xcolor}
\usepackage{tabularx}

\usepackage{xargs}

\usepackage{soul}
\usetikzlibrary{calc}

\makeatletter
\newif\if@anonymize

\@anonymizetrue

\if@anonymize
  \newcommand{\highlight@DoHighlight}{
    \fill [outer sep = -15pt, inner sep = 0pt, color=black]
          ($(begin highlight)+(0,8pt)$) rectangle ($(end highlight)+(0,-3pt)$) ;
  }

  \newcommand{\highlight@BeginHighlight}{
    \coordinate (begin highlight) at (0,0) ;
  }

  \newcommand{\highlight@EndHighlight}{
    \coordinate (end highlight) at (0,0) ;
  }

  \newdimen\highlight@previous
  \newdimen\highlight@current
  \newlength{\item@width}

  \DeclareRobustCommand*\anonymize{%
    \SOUL@setup
    \def\SOUL@preamble{%
      \begin{tikzpicture}[overlay, remember picture]
        \highlight@BeginHighlight
        \highlight@EndHighlight
      \end{tikzpicture}%
    }%
    \def\SOUL@postamble{%
      \begin{tikzpicture}[overlay, remember picture]
        \highlight@EndHighlight
        \highlight@DoHighlight
      \end{tikzpicture}%
    }%
    \def\SOUL@everyhyphen{%
      \discretionary{%
        \SOUL@setkern\SOUL@hyphkern
        \SOUL@sethyphenchar
        \tikz[overlay, remember picture] \highlight@EndHighlight ;%
      }{%
      }{%
        \SOUL@setkern\SOUL@charkern
      }%
    }%
    \def\SOUL@everyexhyphen##1{%
      \SOUL@setkern\SOUL@hyphkern
      \settowidth{\item@width}{##1}%
      \makebox[\item@width]{}%
      \discretionary{%
        \tikz[overlay, remember picture] \highlight@EndHighlight ;%
      }{%
      }{%
        \SOUL@setkern\SOUL@charkern
      }%
    }%
    \def\SOUL@everysyllable{%
      \begin{tikzpicture}[overlay, remember picture]
        \path let \p0 = (begin highlight), \p1 = (0,0) in \pgfextra
          \global\highlight@previous=\y0
          \global\highlight@current =\y1
        \endpgfextra (0,0) ;
        \ifdim\highlight@current < \highlight@previous
          \highlight@DoHighlight
          \highlight@BeginHighlight
        \fi
      \end{tikzpicture}%
      \settowidth{\item@width}{\the\SOUL@syllable}%
      \makebox[\item@width]{}%
      \tikz[overlay, remember picture] \highlight@EndHighlight ;%
    }%
    \SOUL@
  }
\else
  \newcommand{\anonymize}[1]{#1}
\fi
\makeatother

\newcounter{principleno}

\usepackage{xspace}

\definecolor{OwnAzure}{HTML}{336699}
\definecolor{OwnCerulean}{HTML}{CAE2FE}
\definecolor{OwnOliveGreen}{HTML}{556B2F}

\newcommandx{\todoai}[2][1=]{\todo[inline,linecolor=OwnAzure,backgroundcolor=OwnCerulean,bordercolor=OwnAzure,#1]{#2}}

\newcommandx{\addref}[2][1=]
{\todo[inline,linecolor=blue,backgroundcolor=blue!50,bordercolor=blue,#1]{Add reference. #2}}
\newcommandx{\unsure}[2][1=]{\todo[inline, linecolor=red,backgroundcolor=red!25,bordercolor=red,#1]{\textbf{Unsure:} #2}}
\newcommandx{\change}[2][1=]{\todo[inline, linecolor=blue,backgroundcolor=blue!25,bordercolor=blue,#1]{#2}}
\newcommandx{\info}[2][1=]{\todo[linecolor=OwnOliveGreen,backgroundcolor=OwnOliveGreen!25,bordercolor=OwnOliveGreen,#1]{\textbf{Info}: #2}}
\newcommandx{\improvement}[2][1=]{\todo[linecolor=Plum,backgroundcolor=Plum!25,bordercolor=Plum,#1]{\textbf{Improvement}: #2}}
\newcommandx{\thiswillnotshow}[2][1=]{\todo[disable,#1]{#2}}

\newcommand{\para}[1]{\textbf{#1}.}

\usepackage{subfiles}

\newcommand\mytitle{\textcolor{red}{A Review of Serverless Use Cases and their Characteristics}}
\title{Todo title}

\newcommand\TRnumber{Technical Report: SPEC-RG-2021-1\\Version: 1.2}
\newcommand\WGname{SPEC RG Cloud Working Group}

\newcommand\TRdate{\today}
\newcommand\TRcentralURL{research.spec.org}
\newcommand\TRrightURL{www.spec.org}

\newcommand\numAuthors{8}
\makeatletter
\newcommand\defcase[1]{\@namedef{mycase@\the\numexpr#1\relax}}
\newcommand\putAuthors[1]{\@nameuse{mycase@\the\numexpr#1\relax}}
\makeatother

\defcase{0}{
	\begin{textblock}{8}[0,0](\colsinglecentralX,\rowoneY)
		\centering
		\small{\textbf{No Author}}\\
	\end{textblock}	
}

\defcase{1}{
	\begin{textblock}{8}[0,0](\colsinglecentralX,\rowoneY)
		\centering
		\small{\textbf{\authorOneName}}\\
		\footnotesize
		\authorOneAffil
	\end{textblock}	
}
\defcase{2}{
	\begin{textblock}{\authorCellWidth}[0,0](\colDoubleLeftX,\rowoneY)
		\centering
		\small{\textbf{\authorOneName}}\\
		\footnotesize
		\authorOneAffil
	\end{textblock}	
	\begin{textblock}{\authorCellWidth}[0,0](\colDoubleRightX,\rowoneY)
		\centering
		\small{\textbf{\authorTwoName}}\\
		\footnotesize
		\authorTwoAffil
	\end{textblock}	
}
\defcase{3}{
	\begin{textblock}{\authorCellWidth}[0,0](\coloneX,\rowoneY)
		\centering
		\small{\textbf{\authorOneName}}\\
		\footnotesize
		\authorOneAffil
	\end{textblock}	
	\begin{textblock}{\authorCellWidth}[0,0](\coltwoX,\rowoneY)
		\centering
		\small{\textbf{\authorTwoName}}\\
		\footnotesize
		\authorTwoAffil
	\end{textblock}	
	\begin{textblock}{\authorCellWidth}[0,0](\colthreeX,\rowoneY)
		\centering
		\small{\textbf{\authorThreeName}}\\
		\footnotesize
		\authorThreeAffil
	\end{textblock}	
}
\defcase{4}{
	\begin{textblock}{\authorCellWidth}[0,0](\coloneX,\rowoneY)
		\centering
		\small{\textbf{\authorOneName}}\\
		\footnotesize
		\authorOneAffil
	\end{textblock}	
	\begin{textblock}{\authorCellWidth}[0,0](\coltwoX,\rowoneY)
		\centering
		\small{\textbf{\authorTwoName}}\\
		\footnotesize
		\authorTwoAffil
	\end{textblock}	
	\begin{textblock}{\authorCellWidth}[0,0](\colthreeX,\rowoneY)
		\centering
		\small{\textbf{\authorThreeName}}\\
		\footnotesize
		\authorThreeAffil
	\end{textblock}
	\begin{textblock}{8}[0,0](\colsinglecentralX,\rowtwoY)
		\centering
		\small{\textbf{\authorFourName}}\\
		\footnotesize
		\authorFourAffil
	\end{textblock}	
}
\defcase{5}{
	\begin{textblock}{\authorCellWidth}[0,0](\coloneX,\rowoneY)
		\centering
		\small{\textbf{\authorOneName}}\\
		\footnotesize
		\authorOneAffil
	\end{textblock}	
	\begin{textblock}{\authorCellWidth}[0,0](\coltwoX,\rowoneY)
		\centering
		\small{\textbf{\authorTwoName}}\\
		\footnotesize
		\authorTwoAffil
	\end{textblock}	
	\begin{textblock}{\authorCellWidth}[0,0](\colthreeX,\rowoneY)
		\centering
		\small{\textbf{\authorThreeName}}\\
		\footnotesize
		\authorThreeAffil
	\end{textblock}
	\begin{textblock}{\authorCellWidth}[0,0](\colDoubleLeftX,\rowtwoY)
		\centering
		\small{\textbf{\authorFourName}}\\
		\footnotesize
		\authorFourAffil
	\end{textblock}	
	\begin{textblock}{\authorCellWidth}[0,0](\colDoubleRightX,\rowtwoY)
		\centering
		\small{\textbf{\authorFiveName}}\\
		\footnotesize
		\authorFiveAffil
	\end{textblock}	
}
\defcase{6}{
	\begin{textblock}{\authorCellWidth}[0,0](\coloneX,\rowoneY)
		\centering
		\small{\textbf{\authorOneName}}\\
		\footnotesize
		\authorOneAffil
	\end{textblock}	
	\begin{textblock}{\authorCellWidth}[0,0](\coltwoX,\rowoneY)
		\centering
		\small{\textbf{\authorTwoName}}\\
		\footnotesize
		\authorTwoAffil
	\end{textblock}	
	\begin{textblock}{\authorCellWidth}[0,0](\colthreeX,\rowoneY)
		\centering
		\small{\textbf{\authorThreeName}}\\
		\footnotesize
		\authorThreeAffil
	\end{textblock}
	\begin{textblock}{\authorCellWidth}[0,0](\coloneX,\rowtwoY)
		\centering
		\small{\textbf{\authorFourName}}\\
		\footnotesize
		\authorFourAffil
	\end{textblock}	
	\begin{textblock}{\authorCellWidth}[0,0](\coltwoX,\rowtwoY)
		\centering
		\small{\textbf{\authorFiveName}}\\
		\footnotesize
		\authorFiveAffil
	\end{textblock}	
	\begin{textblock}{\authorCellWidth}[0,0](\colthreeX,\rowtwoY)
		\centering
		\small{\textbf{\authorSixName}}\\
		\footnotesize
		\authorSixAffil
	\end{textblock}
}
\defcase{7}{
	\begin{textblock}{\authorCellWidth}[0,0](\coloneX,\rowoneY)
		\centering
		\small{\textbf{\authorOneName}}\\
		\footnotesize
		\authorOneAffil
	\end{textblock}	
	\begin{textblock}{\authorCellWidth}[0,0](\coltwoX,\rowoneY)
		\centering
		\small{\textbf{\authorTwoName}}\\
		\footnotesize
		\authorTwoAffil
	\end{textblock}	
	\begin{textblock}{\authorCellWidth}[0,0](\colthreeX,\rowoneY)
		\centering
		\small{\textbf{\authorThreeName}}\\
		\footnotesize
		\authorThreeAffil
	\end{textblock}
	\begin{textblock}{\authorCellWidth}[0,0](\coloneX,\rowtwoY)
		\centering
		\small{\textbf{\authorFourName}}\\
		\footnotesize
		\authorFourAffil
	\end{textblock}	
	\begin{textblock}{\authorCellWidth}[0,0](\coltwoX,\rowtwoY)
		\centering
		\small{\textbf{\authorFiveName}}\\
		\footnotesize
		\authorFiveAffil
	\end{textblock}	
	\begin{textblock}{\authorCellWidth}[0,0](\colthreeX,\rowtwoY)
		\centering
		\small{\textbf{\authorSixName}}\\
		\footnotesize
		\authorSixAffil
	\end{textblock}
	\begin{textblock}{8}[0,0](\colsinglecentralX,\rowthreeY)
		\centering
		\small{\textbf{\authorSevenName}}\\
		\footnotesize
		\authorSevenAffil
	\end{textblock}
}
\defcase{8}{
	\begin{textblock}{\authorCellWidth}[0,0](\coloneX,\rowoneY)
		\centering
		\small{\textbf{\authorOneName}}\\
		\footnotesize
		\authorOneAffil
	\end{textblock}	
	\begin{textblock}{\authorCellWidth}[0,0](\coltwoX,\rowoneY)
		\centering
		\small{\textbf{\authorTwoName}}\\
		\footnotesize
		\authorTwoAffil
	\end{textblock}	
	\begin{textblock}{\authorCellWidth}[0,0](\colthreeX,\rowoneY)
		\centering
		\small{\textbf{\authorThreeName}}\\
		\footnotesize
		\authorThreeAffil
	\end{textblock}
	\begin{textblock}{\authorCellWidth}[0,0](\coloneX,\rowtwoY)
		\centering
		\small{\textbf{\authorFourName}}\\
		\footnotesize
		\authorFourAffil
	\end{textblock}	
	\begin{textblock}{\authorCellWidth}[0,0](\coltwoX,\rowtwoY)
		\centering
		\small{\textbf{\authorFiveName}}\\
		\footnotesize
		\authorFiveAffil
	\end{textblock}	
	\begin{textblock}{\authorCellWidth}[0,0](\colthreeX,\rowtwoY)
		\centering
		\small{\textbf{\authorSixName}}\\
		\footnotesize
		\authorSixAffil
	\end{textblock}
	\begin{textblock}{\authorCellWidth}[0,0](\colDoubleLeftX,\rowthreeY)
		\centering
		\small{\textbf{\authorSevenName}}\\
		\footnotesize
		\authorSevenAffil
	\end{textblock}
	\begin{textblock}{\authorCellWidth}[0,0](\colDoubleRightX,\rowthreeY)
		\centering
		\small{\textbf{\authorEightName}}\\
		\footnotesize
		\authorEightAffil
	\end{textblock}
}
\defcase{9}{
	\begin{textblock}{\authorCellWidth}[0,0](\coloneX,\rowoneY)
		\centering
		\small{\textbf{\authorOneName}}\\
		\footnotesize
		\authorOneAffil
	\end{textblock}	
	\begin{textblock}{\authorCellWidth}[0,0](\coltwoX,\rowoneY)
		\centering
		\small{\textbf{\authorTwoName}}\\
		\footnotesize
		\authorTwoAffil
	\end{textblock}	
	\begin{textblock}{\authorCellWidth}[0,0](\colthreeX,\rowoneY)
		\centering
		\small{\textbf{\authorThreeName}}\\
		\footnotesize
		\authorThreeAffil
	\end{textblock}
	\begin{textblock}{\authorCellWidth}[0,0](\coloneX,\rowtwoY)
		\centering
		\small{\textbf{\authorFourName}}\\
		\footnotesize
		\authorFourAffil
	\end{textblock}	
	\begin{textblock}{\authorCellWidth}[0,0](\coltwoX,\rowtwoY)
		\centering
		\small{\textbf{\authorFiveName}}\\
		\footnotesize
		\authorFiveAffil
	\end{textblock}	
	\begin{textblock}{\authorCellWidth}[0,0](\colthreeX,\rowtwoY)
		\centering
		\small{\textbf{\authorSixName}}\\
		\footnotesize
		\authorSixAffil
	\end{textblock}
	\begin{textblock}{\authorCellWidth}[0,0](\coloneX,\rowthreeY)
		\centering
		\small{\textbf{\authorSevenName}}\\
		\footnotesize
		\authorSevenAffil
	\end{textblock}	
	\begin{textblock}{\authorCellWidth}[0,0](\coltwoX,\rowthreeY)
		\centering
		\small{\textbf{\authorEightName}}\\
		\footnotesize
		\authorEightAffil
	\end{textblock}	
	\begin{textblock}{\authorCellWidth}[0,0](\colthreeX,\rowthreeY)
		\centering
		\small{\textbf{\authorNineName}}\\
		\footnotesize
		\authorNineAffil
	\end{textblock}
}
\defcase{10}{
	\begin{textblock}{\authorCellWidth}[0,0](\coloneX,\rowoneY)
		\centering
		\small{\textbf{\authorOneName}}\\
		\footnotesize
		\authorOneAffil
	\end{textblock}	
	\begin{textblock}{\authorCellWidth}[0,0](\coltwoX,\rowoneY)
		\centering
		\small{\textbf{\authorTwoName}}\\
		\footnotesize
		\authorTwoAffil
	\end{textblock}	
	\begin{textblock}{\authorCellWidth}[0,0](\colthreeX,\rowoneY)
		\centering
		\small{\textbf{\authorThreeName}}\\
		\footnotesize
		\authorThreeAffil
	\end{textblock}
	\begin{textblock}{\authorCellWidth}[0,0](\coloneX,\rowtwoY)
		\centering
		\small{\textbf{\authorFourName}}\\
		\footnotesize
		\authorFourAffil
	\end{textblock}	
	\begin{textblock}{\authorCellWidth}[0,0](\coltwoX,\rowtwoY)
		\centering
		\small{\textbf{\authorFiveName}}\\
		\footnotesize
		\authorFiveAffil
	\end{textblock}	
	\begin{textblock}{\authorCellWidth}[0,0](\colthreeX,\rowtwoY)
		\centering
		\small{\textbf{\authorSixName}}\\
		\footnotesize
		\authorSixAffil
	\end{textblock}
	\begin{textblock}{\authorCellWidth}[0,0](\coloneX,\rowthreeY)
		\centering
		\small{\textbf{\authorSevenName}}\\
		\footnotesize
		\authorSevenAffil
	\end{textblock}	
	\begin{textblock}{\authorCellWidth}[0,0](\coltwoX,\rowthreeY)
		\centering
		\small{\textbf{\authorEightName}}\\
		\footnotesize
		\authorEightAffil
	\end{textblock}	
	\begin{textblock}{\authorCellWidth}[0,0](\colthreeX,\rowthreeY)
		\centering
		\small{\textbf{\authorNineName}}\\
		\footnotesize
		\authorNineAffil
	\end{textblock}
	\begin{textblock}{8}[0,0](\colsinglecentralX,\rowfourY)
		\centering
		\small{\textbf{\authorTenName}}\\
		\footnotesize
		\authorTenAffil
	\end{textblock}
}

\newcommand\authorOneName{Simon Eismann}
\newcommand\authorOneAffil{
	Julius-Maximilian University \\
	Würzburg\\ Germany\\
	\emph{simon.eismann@uni-wuerzburg.de}
}

\newcommand\authorTwoName{Joel Scheuner}
\newcommand\authorTwoAffil{
	Chalmers | University of Gothenburg \\
	Gothenburg\\
	Sweden\\
	\emph{scheuner@chalmers.se}
}

\newcommand\authorThreeName{Erwin van Eyk}
\newcommand\authorThreeAffil{
Vrije Universiteit\\ Amsterdam\\ The
Netherlands
	\emph{e.vaneyk@atlarge-research.com}
}

\newcommand\authorFourName{Maximilian Schwinger}
\newcommand\authorFourAffil{
	German Aerospace Center \\
	Oberpfaffenhofen\\
	Germany\\
	\emph{maximilian.schwinger@dlr.de}
}

\newcommand\authorFiveName{Johannes Grohmann}
\newcommand\authorFiveAffil{
	Julius-Maximilian University \\
	Würzburg\\
	Germany\\
	\emph{johannes.grohmann@uni-wuerzburg.de}
}

\newcommand\authorSixName{Nikolas Herbst}
\newcommand\authorSixAffil{
	Julius-Maximilian University \\
	Würzburg\\
	Germany\\
	\emph{nikolas.herbst@uni-wuerzburg.de}
}

\newcommand\authorSevenName{Cristina L. Abad}
\newcommand\authorSevenAffil{
    Escuela Superior Politecnica del Litoral \\ Guayaquil \\ Ecuador \\
	\emph{cabad@fiec.espol.edu.ec}
}

\newcommand\authorEightName{Alexandru Iosup}
\newcommand\authorEightAffil{
Vrije Universiteit\\ Amsterdam\\ The
Netherlands
	\emph{A.Iosup@atlarge-research.com}
}
\newcommand\authorNineName{}
\newcommand\authorNineAffil{
	\emph{}
}

\newcommand\authorTenName{}
\newcommand\authorTenAffil{

	\emph{}
}

\begin{document}
 
\selectlanguage{english} 
\frontmatter

%% titlepage.tex
%%
\thispagestyle{empty}
% coordinates for the bg shape on the titlepage
\newcommand{\changefont}[3]{\fontfamily{#1} \fontseries{#2} \fontshape{#3} \selectfont}
\newcommand{\diameter}{20}
\newcommand{\xone}{-25}
\newcommand{\xtwo}{165}
\newcommand{\yone}{20}
\newcommand{\ytwo}{-253}

%%%%%%%%%%%%%%%%%%%%%%%%%
%%%%% Authors array %%%%%
%%%%%%%%%%%%%%%%%%%%%%%%%
\newcommand{\rowoneY}{5.5}		
\newcommand{\rowtwoY}{7.0}
\newcommand{\rowthreeY}{8.5}
\newcommand{\rowfourY}{10.1}

% three authors in line
\newcommand{\coloneX}{2.5}
\newcommand{\coltwoX}{7.45}
\newcommand{\colthreeX}{12.4}

% two authors in line
\newcommand{\colDoubleLeftX}{5}
\newcommand{\colDoubleRightX}{10}

% single author in line
\newcommand{\colsinglecentralX}{5.9}

% width of a box for a single author cell
\newcommand{\authorCellWidth}{4.9}

\begin{titlepage}
% Frame shape
\begin{tikzpicture}[overlay]
\draw[color=gray]  
 (\xone mm, \yone mm) -- (\xtwo mm, \yone mm) arc (90:0:\diameter pt) 
  -- (\xtwo mm + \diameter pt , \ytwo mm) -- (\xone mm + \diameter pt , \ytwo mm) 
 arc (270:180:\diameter pt) -- (\xone mm, \yone mm);
\end{tikzpicture}

\changefont{phv}{m}{n}	% helvetica	

%%%%%%%%%%%%%%%%%%%%%%%%%%%%%%% Title
\begin{textblock}{14}[0,0](3,2.3)
	\centering
	\large{\TRnumber}\\
	\vspace*{1cm}
	\huge{\mytitle}\\
	\vspace*{0.5cm}
	\Large{\WGname}
\end{textblock}
%%%%%%%%%%%%%%%%%%%%%%%%%%%%%%% Red bar
\begin{textblock}{15.5}[0,0](2,5.2)
	\begin{tikzpicture}
		\fill[red!80!brown] (0,0cm) rectangle (19.5cm,0.1cm);
	\end{tikzpicture}
\end{textblock}

%%%%%%%%%%%%%%%%%%%%%%%%%%%%%%%%%%%%%%%%%%%%%%%%%%%%%%%%%%%%%%
%%%%%%%%%%%%%%%%%%%%%%%%%%%%%%% Authors %%%%%%%%%%%%%%%%%%%%%%
%%%%%%%%%%%%%%%%%%%%%%%%%%%%%%%%%%%%%%%%%%%%%%%%%%%%%%%%%%%%%%
\begin{center}
	\putAuthors{\numAuthors}
\end{center}

%%%%%%%%%%%%%%%%%%%%%%%%%%%%%%% Logo
\begin{textblock}{14}[0,0](3,13)
	\hfill
	\includegraphics[width=3cm]{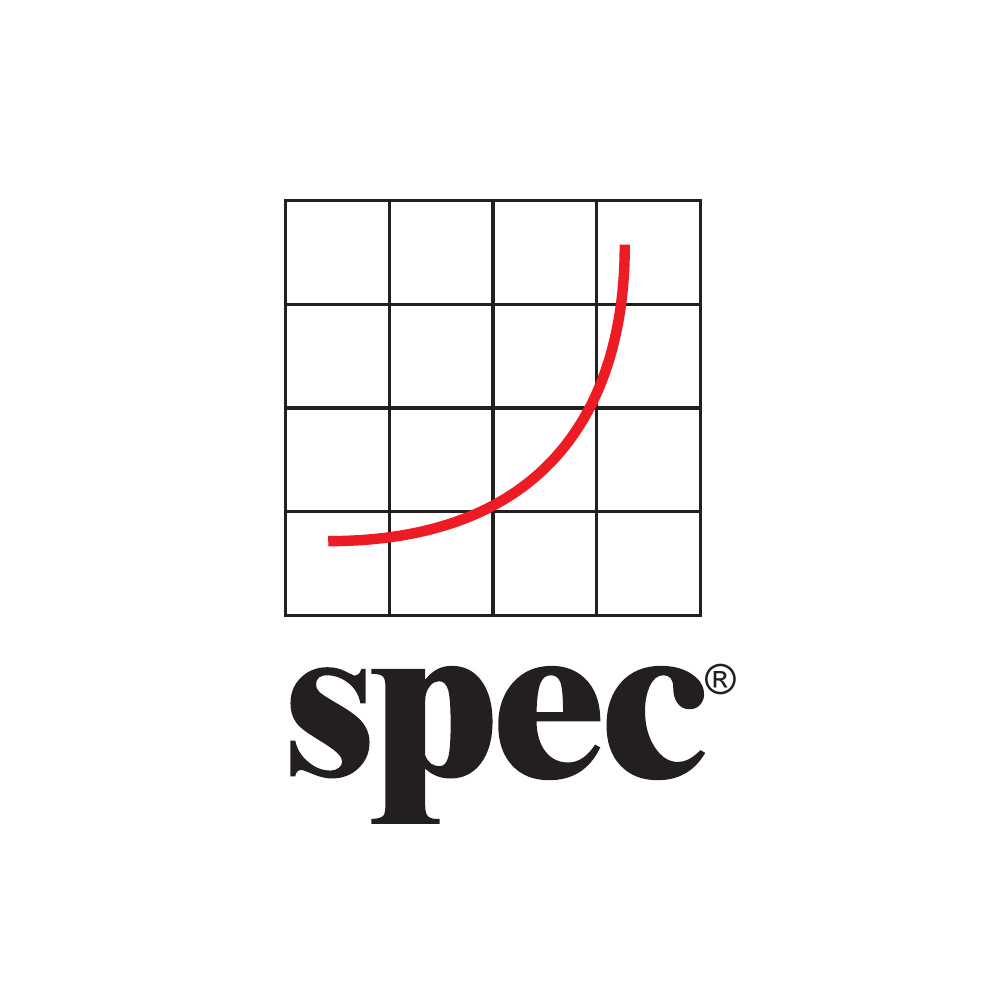} \hfill
	\includegraphics[width=1.9cm]{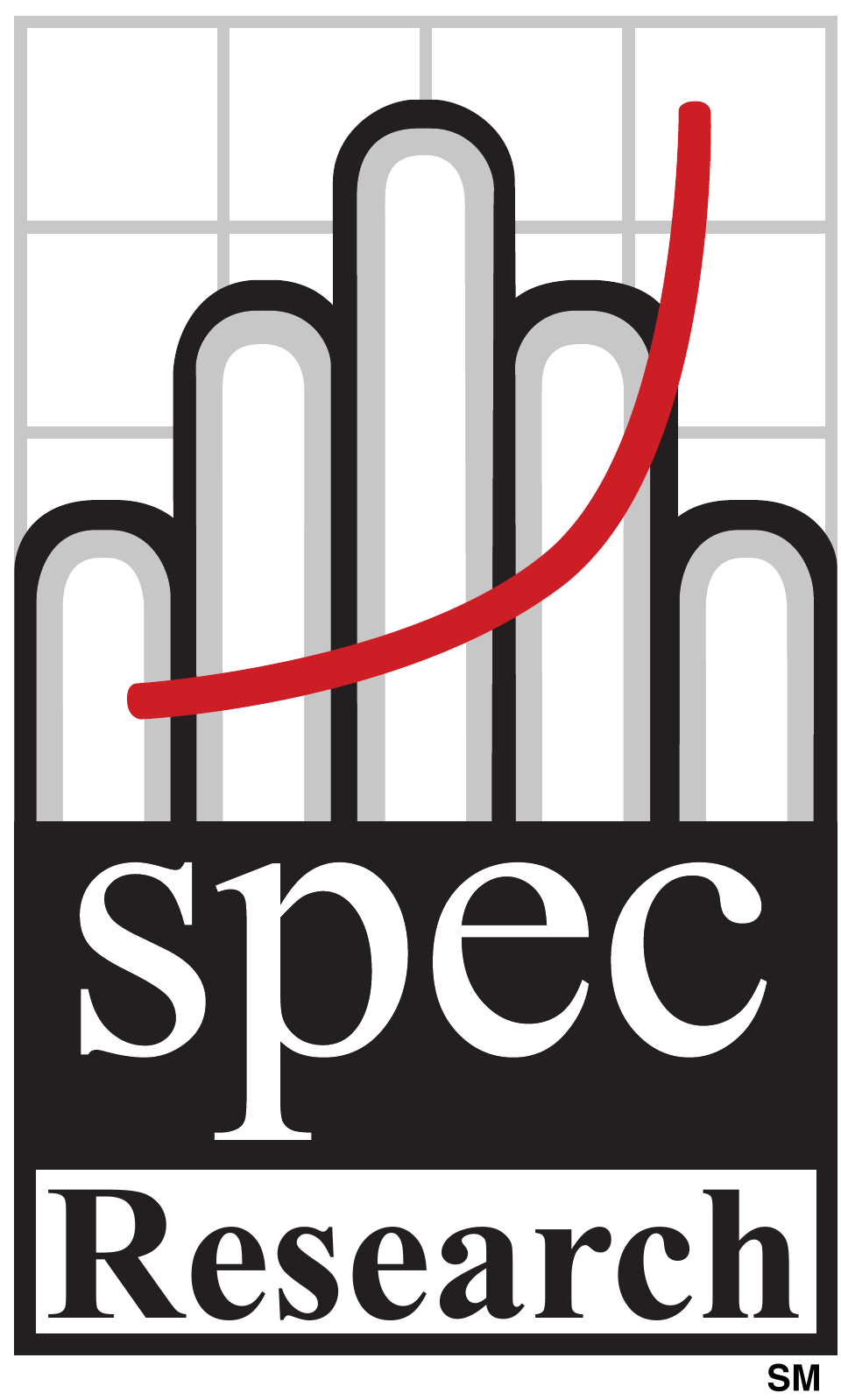} \hspace{1.1cm}\hfill
	\includegraphics[width=1.9cm]{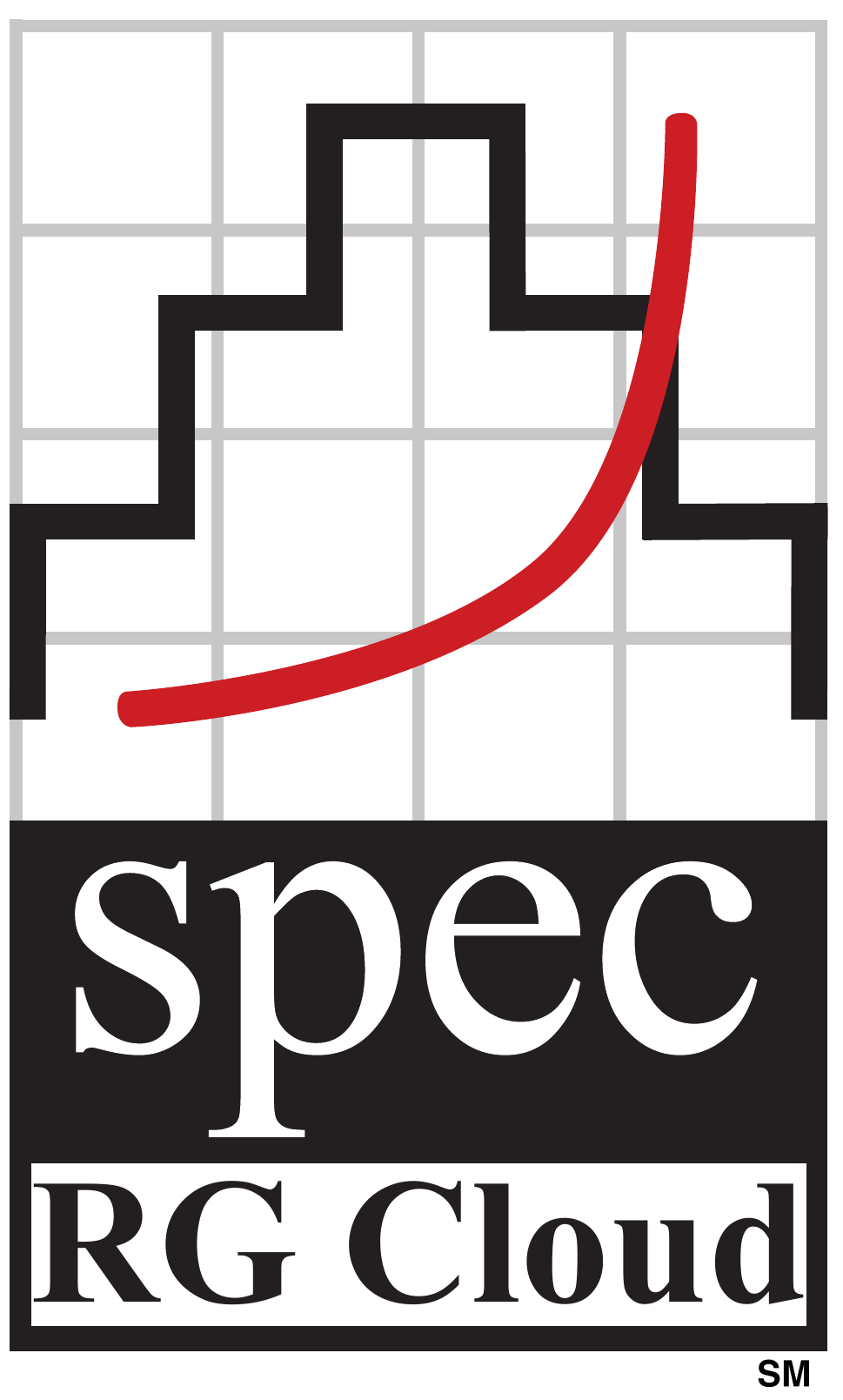} \hspace{1.1cm}\hfill
	\hfill
\end{textblock}

%%%%%%%%%%%%%%%%%%%%%%%%%%%%%%% Acknowledgements

%%%%%%%%%%%%%%%%%%%%%%%%%%%%%%% Footer
% below the frame
\begin{textblock}{14}[0,0](3,16.75)
	\centering
	\large{\textbf{\TRdate}}
	\hfill
	\large{\textbf{\TRcentralURL}}
	\hfill
	\large{\textbf{\TRrightURL}}
\end{textblock}

\end{titlepage}

\newpage 
\section*{Document History}
\begin{itemize}
    \item[\textbf{v1.0}] Initial release
    \item[\textbf{v1.1}] Added discussion of conflicting report to the introduction\\ Added discussion of cost pitfalls to Section~\ref{sec:results:requirements:motivation}
    \item[\textbf{v1.2}] Added new data for the application type (Section~\ref{subsec:application_type})
\end{itemize}

\newpage
\pagenumbering{roman}
\setcounter{tocdepth}{4}
\begin{spacing}{1.3}
\tableofcontents
\end{spacing}

\newpage
\thispagestyle{plain}
\section*{Executive Summary}
The serverless computing paradigm promises many desirable properties for cloud applications---low-cost, fine-grained deployment, and management-free operation.
Consequently, the paradigm has underwent rapid growth: there currently exist tens of serverless platforms and all global cloud providers host serverless operations.
To help tune existing platforms, guide the design of new serverless approaches, and overall contribute to understanding this paradigm, in this work we present a long-term, comprehensive effort to identify, collect, and characterize serverless use cases.
We survey 89 use cases, sourced from white and grey literature, and from consultations with experts in areas such as scientific computing.
We study each use case using 24 characteristics, including general aspects, but also workload, application, and requirements. When the use cases employ workflows, we further analyze their characteristics.
Overall, we hope our study will be useful for both academia and industry, and encourage the community to further share and communicate their use cases.
\\

\noindent\textbf{Keywords}\\

\noindent serverless use cases, cloud computing, serverless computing, serverless applications, workflows, requirements analysis, empirical research\\

\noindent\textbf{Disclaimer}\\

\noindent SPEC, the SPEC logo and the names SPEC CPU2017, SERT, SPEC Cloud IaaS 2018 are trademarks of the Standard Performance Evaluation Corporation (SPEC). 
SPEC Research and SPEC RG Cloud are service marks of SPEC. Additional product and service names mentioned herein may be the trademarks of their respective owners. 
Copyright © 1988-2020 Standard Performance Evaluation Corporation (SPEC). 
All rights reserved.

\mainmatter
\setlength{\parindent}{0cm}

\section{Introduction}\label{sec:intro}\label{sec:introduction}

Serverless computing is an emerging technology with increasing impact on our modern society, and increasing adoption by both academia and industry~\cite{FutureScape, Markets, 10.5555/3027041.3027047}.
The key promise of serverless computing is to make computing services more accessible, fine-grained, and affordable~\cite{DBLP:journals/internet/EykTTVUI18,DBLP:journals/corr/abs-1902-03383} by managing operational concerns~\cite{10.1145/3368454}.
Major cloud providers, such as Amazon, Microsoft, Google, and IBM already offer capable serverless platforms.
However, serverless computing, and its common Function-as-a-Service~(FaaS) realization, still raises many important challenges that may reduce adoption. These challenges have been recognized and discussed in fields such as software engineering, distributed systems, performance engineering~\cite{DBLP:conf/middleware/EykIST17,DBLP:conf/wosp/EykIAGE18,DBLP:conf/cidr/HellersteinFGSS19}.
This work focuses on a first step to alleviate these challenges:
{\it understanding serverless applications through a variety of use cases}.
\newline

Serverless computing enables developers to focus on implementing business logic, leaving the operational concerns to cloud providers. In turn, the providers turn to automation, which they achieve through 
capable serverless platforms, such as AWS Lambda, Azure Functions, or Google Cloud Functions, and IBM Cloud Functions (based on Apache OpenWhisk).
Serverless platforms already support fine-grained function deployment, detailed resource allocation, and to some extent also autoscaling~\cite{10.1145/3368454}.
However, more sophisticated operational features have started to emerge such as:
(a) complex function composition and even full {\it workflows}, 
(b) eventing and provider-managed messaging, 
(c) low-latency scheduling, 
(d) file storage and database setup, 
(e) streaming and locality-aware deployment, and 
(f) versioning and logging solutions. 
These features facilitate the serverless {\it application lifecyle}, and help further decreases the time-to-market for serverless applications~\cite{leitner2019MixedMethod}.
\newline

Researchers and industry practitioners have an urgent need for serverless use cases.
The variety of already existing platforms and support from the major cloud providers indicate the presence of many serverless applications.
However, relatively little is known about their characteristics or behavior.
For an emerging technology such as serverless computing, researchers, 
engineers, and 
platform providers 
could use descriptions of {\it use cases}---{\it which applications?}, {\it where and how was this technology already successfully applied?}, and {\it what are the characteristics of these use cases?}---to guide their drive for discovery and improvement {\it in the right direction}.
\textit{Researchers} can study different use cases related to the same application to extract meaningful patterns and trigger new designs. They can also identify representative use cases, which can later be used for the evaluation of novel approaches and in empirical studies. 
\textit{Engineers} require descriptions in which areas serverless computing was already successfully applied, which helps to decide whether to adopt serverless computing for other projects. Additionally, existing solutions can serve as blueprints for similar use cases. \textit{Platform providers} require knowledge of how their products are used, to optimize them and gaps in adoption can point out deficits in their current offerings.
\newline

There are only a few, and sometimes conflicting, reports addressing important questions such as why developers build serverless applications, when serverless applications are well suited, or how serverless applications are implemented in practice. For example, there are reports of significant cost savings by switching to serverless applications~\cite{adzic2017serverless, Levinson2020}, but also articles suggesting higher cost in some scenarios compared to traditional hosting~\cite{eivy2017wary}. There are also reports of successfully serverless applications for data-intensive applications~\cite{witte2019serverless, 10.1145/3307339.3343462}, despite other articles claiming that serverless is not well suited for data-intensive applications~\cite{DBLP:conf/cidr/HellersteinFGSS19}. Some people suggest that containers are superior to serverless for latency-critical tasks~\cite{Thorn}, but there are also reports of people successfully applying serverless for latency-critical user-facing traffic~\cite{Droplr}. Having concrete information on these topics would be valuable for managers to guide decisions on whether a serverless application can be a suitable solution for a specific use case.
\newline

However much needed, serverless use cases have not been studied systematically so far.
For serverless computing, existing research has focused on serverless platforms and their performance properties~\cite{yussupov:19}.
Several studies currently exist about the features, architecture, and performance properties of these platforms~\cite{DBLP:journals/internet/EykIGEBVTSHA19,10.1007/978-3-319-99819-0_11, figiela2018performance, lee2018evaluation, lloyd2018serverless, wang2018peeking}.
Shahrad et al.~\cite{shahrad2020serverless} characterize the aggregated performance properties of the entire production FaaS workload from Microsoft Azure Functions, but do not provide details on individual use cases.
A recent mixed-method empirical study investigates how developers use serverless computing, focusing on the issues (pain points) they experienced~\cite{leitner2019MixedMethod}.
Another multivocal literature review discusses common patterns in the architecture of serverless applications~\cite{taibi2020serverless}.
To the best of our knowledge, the only existing collection of serverless use cases is an article by Castro et al.~\cite{10.1145/3368454}, which introduces ten use cases collected from non-peer-reviewed (\textit{grey}) literature.
\newline

In this technical report, we collect a total of \totalusecases serverless use cases from four different sources. \openusecases use cases are from open-source projects, \whiteusecases from white literature, \greyusecases from grey literature, and \scientificusecases from the area of scientific computing. Each use case is reviewed by a pair of reviewers in regard to 24 characteristics, such as execution pattern, workflow coordination, use of external services, and motivation for adopting serverless.
The full dataset containing all use cases and their characteristics is publicly available as a persistent Zenodo repository~\cite{dataset}. 
\newline

In the next section, we discuss our process for use-case collection and characterization. In Section~\ref{sec:results}, we describe the \characteristics characteristics we reviewed for each use case and the results of this review. Section~\ref{sec:validity} discusses threats to validity and mitigation strategies.
Finally, Section~\ref{sec:conclusion} concludes this technical report, and discusses promising future research directions based on the finding of this study.

\section{Study Design}
\label{sec:method}

This section summarizes our overall study process, describes the data sources to identify primary studies, the selection strategy with inclusion and exclusion criteria, the characteristics review protocol, and the discussion and consolidation phase covering inter-reviewer agreement.

\subsection{Process Overview}
\label{sec:process}

\begin{figure}[ht]
    \centering
    \includegraphics[width=\linewidth]{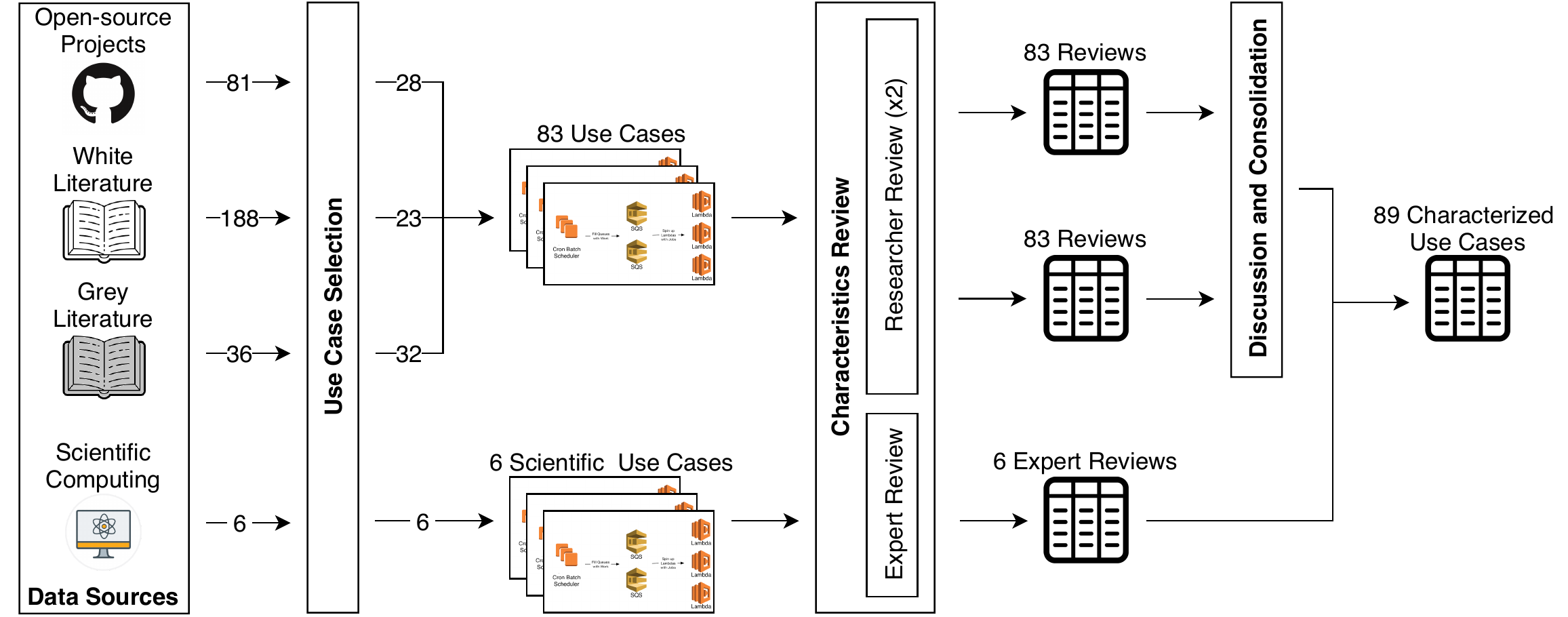}
    \caption{Process overview for use case analysis.}
    \label{fig:overview}
\end{figure}

\Cref{fig:overview} summarizes the use case analysis process.
Firstly, we compiled an extensive list of potentially relevant use cases from four different data sources (see \Cref{sec:process:source}), namely open source projects, white literature, grey literature, and scientific computing.
Secondly, we applied our selection criteria (see \Cref{subsec:selection}) to classify and filter only relevant use cases in the context of this study.
This resulted in 83 use cases from publicly available sources and \scientificusecases scientific use cases from internal sources, where we had access to domain experts.
Thirdly, we defined a list of interesting characteristics including potential values and perform reviews to extract the actual values from available documentation  (see \Cref{subsec:data_extraction}).
For all public sources, 2 randomly assigned researchers out of a pool of 7 available authors conducted two redundant reviews for each use case.
Each scientific use case was reviewed by a single domain expert.
Subsequently, we calculated the inter-reviewer agreements for all redundant reviews and resolved any conflicting values during discussion and consolidation (see \Cref{subsec:data_synthesis}).
This resulted in a total of \totalusecases analyzed use cases.

\subsection{Data Sources}
\label{sec:process:identification}\label{sec:process:source}\label{sec:process:sourcing}

Reports on use cases for serverless applications appear in many different forms ranging from peer-reviewed academic papers, open-source projects, blog posts, podcasts, talks, provider-reported success stories to direct exchange with application developers. Therefore, we collect use cases from a variety of different sources. We also aim to not have a dominant source that contributes the lions share of the use cases and therefore introduces a strong selection bias. Based on this, we do not aim for an exhaustive collection of use cases, but collect use cases from the following different sources with the goal of obtaining a large varied sample:

\begin{itemize}
    \item \textbf{Open-source projects:} Many serverless open-source projects are currently available on GitHub.
    As a starting point for the open-source projects, we used an existing data set from \cite{pavlov:19}. This data set was scraped from GitHub using GHTorrent~\cite{Gousi13}, an offline mirror of the  GitHub public event time line. It excludes unrelated or insignificant projects, based on a keyword search\footnote{Keywords: \textit{aws, aws lambda, amazon lambda, lambda functions, azure, openwhisk, serverless, google cloud functions, microsoft azure, azure functions, ibm blue mix, bluemix, oracle fn, oracle cloud fn, kubeless, ibm cloud functions, fn project}} and also excludes any projects that started prior to the launch of AWS Lambda (the first major serverless platform). From this data set, we removed small and inactive projects based on the number of files, commits, contributors, and watchers. As this still left us with many projects that only mention one of the keywords (e.g., ''In the future, we are looking to use AWS lambda for the image resizing.''), we manually filtered the resulting data set to include only projects that are deployed as serverless applications. This resulted in a total of \openusecases use cases from open-source projects.
    
    \item \textbf{White literature:} There is also a growing interest in serverless applications from academia, which results in a number of scientific publications, i.e., journal papers, conference papers and workshop papers describing serverless use cases. For white literature, we based our search on an existing community-curated dataset on literature for serverless computing consisting of over 180 articles from 2016 to 2019~\cite{sldataset}. First, we filtered the articles based on title and abstract. In a second iteration, we filtered out any articles that implement only a single function for evaluation purposes or do not include sufficient detail to enable a review. As the authors were familiar with a few additional publications describing serverless applications, we contributed these articles to the community-curated dataset and included them in this study. This resulted in a total of \whiteusecases use cases from white literature.
    
    \item \textbf{Grey literature:} In  software engineering, the discourse in not limited to scientific articles but extends to grey literature, such as blog posts, forum discussions and podcasts~\cite{10.1145/2915970.2916008, 10.1145/2786805.2803200}. For serverless computing, there are a number of blog posts by companies or individuals, talks at industry conferences and provider-reported success stories, as the development of serverless computing was initially mostly industry driven. We filtered the case studies reported by the major serverless providers (AWS\footnote{\url{https://aws.amazon.com/solutions/case-studies/}}, Azure\footnote{\url{https://azure.microsoft.com/en-in/case-studies/}}, Google\footnote{\url{https://cloud.google.com/customers}} and IBM\footnote{\url{https://www.ibm.com/case-studies/}}) and selected those that used mostly serverless solutions. We also included the ten use cases reported in a recent article on the rise of serverless computing~\cite{10.1145/3368454}, which to the best of our knowledge is the largest collection of grey literature on serverless use cases. We further extended this collection with grey literature articles describing serverless use cases that the authors were already familiar with. This process resulted in a total of \greyusecases use cases from grey literature.
    
    \item \textbf{Scientific computing:} There is also an increasing interest in serverless solutions from the scientific computing community (e.g., by NASA~\cite{nasa}). However, most of these use cases are still at an early stage and therefore there is little public data available for them. One of the authors of this paper is currently employed at the German Aerospace Center~(DLR), which allowed us to collect information about several projects at DLR that are either currently moving to serverless solutions or are planning to do so. Additionally, a use case from the German Electron Syncrotron~(DESY) could be included. This resulted in a total of \scientificusecases use cases from the area of scientific computing.
\end{itemize}
Some use cases are contained in multiple sources, e.g., a use case might have a GitHub repository that matches our keywords and is also used in the evaluation of an academic paper. For these use cases, we assign them only to a single source using the following ranking: open-source projects > grey literature > white literature. For the scientific use cases, there are no overlaps with the other use case sources.

\subsection{Use Case Selection}
\label{subsec:selection}

We defined the following inclusion (I) and exclusion (E) criteria for our study:
\begin{itemize}
    \item[I1] Concrete serverless use cases, as we are interested in real-world example applications.
    \item[I2] Use cases described in sufficient detail to conduct a meaningful review (i.e., excluding vague high-level case studies mainly focusing on a specific serverless platform or solution, but lacking technical detail).
    \item[E1] Serverless platforms (e.g., Apache OpenWhisk) and frameworks (e.g., Serverless Framework\footnote{\url{https://www.serverless.com/}}), as these are not concrete workloads.
    \item[E2] Boilerplate code and simple technology demonstrations as often found in official serverless provider documentation, as these do not constitute full-fledged use cases.
    \item[E3] Academic papers on the same use case. For example, there are a number of academic papers that discuss serverless neural networks serving~\cite{ishakian2018serving, bhattacharjee2019barista, tu2018pay}. In this case we only include a single representative paper.
\end{itemize}

\subsection{Characteristics Review}
\label{subsec:data_extraction}
We first determined and formalized the set of investigated characteristics. In an initial round, all authors individually suggested characteristics they consider interesting. In a next round, we merged similar characteristics and kept all characteristics that at least two authors considered relevant. This process resulted in 24 characteristics, which can be divided into five groups: general characteristics, workload characteristics, application characteristics, requirement characteristics, and workflow characteristics. \emph{General characteristics} aim to quantify the structure of our data set and include characteristics such as ``Is the use case open-source?'', ``Is the use case currently deployed in production?'', and ``What domain is this use case from?''. \emph{Workload characteristics} aim to describe the traffic pattern and request properties of the use case, e.g., ``Is the workload bursty?'', ``What is the data volume per request?'', and ``Is the application workload triggered by HTTP requests, cloud events, or regularly scheduled?''. \emph{Application characteristics} describe the structure and properties of the serverless application itself and focuses on characteristics such as ``How many functions does the application consist of?'', ``What programming languages are used?'', and ``Which managed cloud services does the application use?''. \emph{Requirement characteristics} describe the requirements from the stakeholders, such as ``Is latency relevant?'', ``Does the application have to run in a specific region, replicated in multiple regions or even on edge devices?'', and ``What is the reason for the adoption of serverless computing?''. Finally, \emph{workflow characteristics} describe the properties of workflows within the serverless applications, e.g., ``Is the use case a workflow?'', ``How many functions does the workflow consist of?'', and ``How is the workflow execution coordinated?''. \\

Based on a group discussion, we defined an exhaustive set of potential values for each characteristics. For example, for the characteristic ``How are executions triggered?'' we defined the potential values ``HTTP request'', ``Cloud event', ``Scheduled'' and ``Manual''. Additionally, for every characteristic we introduced the values ``Unknown'' and ``Not applicable''. ``Unknown'' indicates that the documentation of the use case does not contain enough information to determine this characteristic. ``Not applicable'' is used when a characteristic does not make sense for a use cases, for example all workflow characteristics are only applicable to use cases that contain a workflow. For some characteristics, we were not able to define a set of potential values prior to reviewing the use cases. For these characteristics, we used text fragments during the review. Using thematic coding~\cite{coffey1996making, guest2011applied}, we extracted codes and treated those as the values for these characteristics. For example, for the characteristic ``What is the reason for the adoption of serverless computing?'' thematic coding resulted in the codes ``NoOps'', ``Scalability'', ``Performance'', ``Maintainability'' and ``Simplify Development''. This process enabled us to extract quantifiable results from the textual descriptions. \\

We randomly assigned each use case to 2 reviewers out of a pool of 7 available reviewers from the authors.
We manually adjusted a few reviewer assignments to minimize the number of coinciding reviewer pairs (i.e., avoid that many use cases are reviewed by the same two reviewers).
Subsequently, each reviewer individually assigned values to all characteristics of its assigned use cases.

\subsection{Discussion and Consolidation}
\label{subsec:data_synthesis}
After completing the initial round of reviews, we calculate the fleiss kappa to quantify the level of agreement between the reviewers~\cite{gwet2014handbook}. Due to the nature of the fleiss kappa, we excluded all characteristic assignments, where at least one reviewer assigned multiple values for a characteristic (e.g., if a use case execution is triggered both via HTTP requests and cloud events), the characteristics using thematic coding as well as the numeric characteristic ``How many functions does the application consist of?''. As characteristics have a different number of possible values, we calculated an individual fleiss kappa value for each characteristic and then the weighted average across these individual kappa fleiss values. This results in a kappa fleiss value of 0.48, which can be interpreted as ``moderate agreement''~\cite{landis1977measurement}.

In the following discussion and consolidation phase, the reviewers compared their notes and tried to reach a consensus for the characteristics with conflicting assignments. In a few cases, the two reviewers had different interpretations of a characteristics. These conflicts were discussed among all authors to ensure that characteristic interpretations were consistent. However for most conflicts, the consolidation was a quick process as the most frequent type of conflict was that one reviewer found additional documentation that the other reviewer did not find.
Following this process, we were able to resolve all conflicts, resulting in a collection of \totalusecases use cases described by 24 characteristics. \\

For the scientific use cases, a different approach was necessary as many of them were not publicly available yet. Therefore, these use cases are reviewed by a single domain expert, which is either involved in the development of the use case or in direct contact with the development. For each of the scientific use cases there is also a textual description (see Appendix Section~\ref{appendix}).

\section{Analysis Results: On the Characteristics of Serverless Use Cases}
\label{sec:results}

We describe in this section the results of our characterization and analysis of serverless use cases. Overall, we cover a diverse set of  characteristics, identifying the values commonly used in practice and further analyzing their impact on serverless practice.

\subsection{Main Findings}
\label{sec:results:main}

Our main findings are:
\begin{enumerate}

    \item \emph{General Characteristics:} We find AWS as the currently dominating for platform for serverless applications (80\%). The dominating application domain is web services (33\%), with 40\% of the analysed workloads being business-critical and at least 55\% of them in production already.
    
    \item \emph{Application Characteristics:}
    82\% of all use cases consist of applications that use five or less different functions. Most (67\%) of these functions are short-running, with running times in the order of milliseconds or seconds. JavaScript and Python are the most used programming languages for cloud functions (each used by 32\% of the cases we studied).
    These applications depend on a wide variety of cloud services, with the three most used ones being cloud storage (used by 61\% of the applications) and cloud database (47\%); cloud API gateway (18\%) and cloud pub-sub (17\%) are also widely used.
    
    \item \emph{Requirements Characteristics:} The reduced operation cost of serverless platforms (33\%), the reduced operation effort (24\%), the scalability (24\%), and performance gains (13\%) are the main drivers of serverless adoption. 
    In comparison, cost savings seems to be a stronger motivator than the performance benefits. 
    At the same time, 58\% of use cases have latency requirements, 2\% even have real-time demands, while only 36\% are latency insensitive.
    Locality requirements are only relevant for 21\% of the total use cases.
    
    \item \emph{Workload Characteristics:}
    81\% of the analyzed use cases exhibit bursty workloads.
    This highlights the overall trend of serverless workloads to feature unpredictable on-demand workloads, typically triggered through lightweight (<1MB) HTTP requests.

    \item \emph{Workflow Characteristics:}
    Although the presence of workflows is already sizable (31\% of the use cases), most workflows are of simple structure, small, and short-lived. 
    This is likely to change, as demand follows natural trends and orchestration methods move toward (cloud-native) workflow engines.
    
\end{enumerate}

\subsection{General Characteristics}
\label{sec:results:general}

In this section, we analyse general characteristics of serverless use cases: the supported platform(s) and application types. Furthermore, we check if a serverless use case is in production yet and its availability as open source. Last, we report on the distribution across application domains for the analysed use cases.

\subsubsection{Platform}
\para{Description}
In November 2014, Amazon released the first commercial Function-as-a-Service platform with AWS Lambda and started the serverless trend. Two years later in 2016, Microsoft Azure, Google Cloud, and IBM Cloud released their own Function-as-a-Service platforms. There are also a number of open-source Function-as-a-Service platforms, such as Knative, OpenWhisk and OpenLambda. Selecting a deployment platform is a major decision for serverless applications, as there is a strong vendor lock-in that makes changing the deployment platform at a later point in time difficult. In this study, we grouped the deployment platforms into \emph{AWS}, \emph{Azure}, \emph{IBM Cloud}, \emph{Google Cloud}, and \emph{Private Cloud}. \\

\begin{figure}[h]
    \centering
    \includegraphics[width=0.8\linewidth]{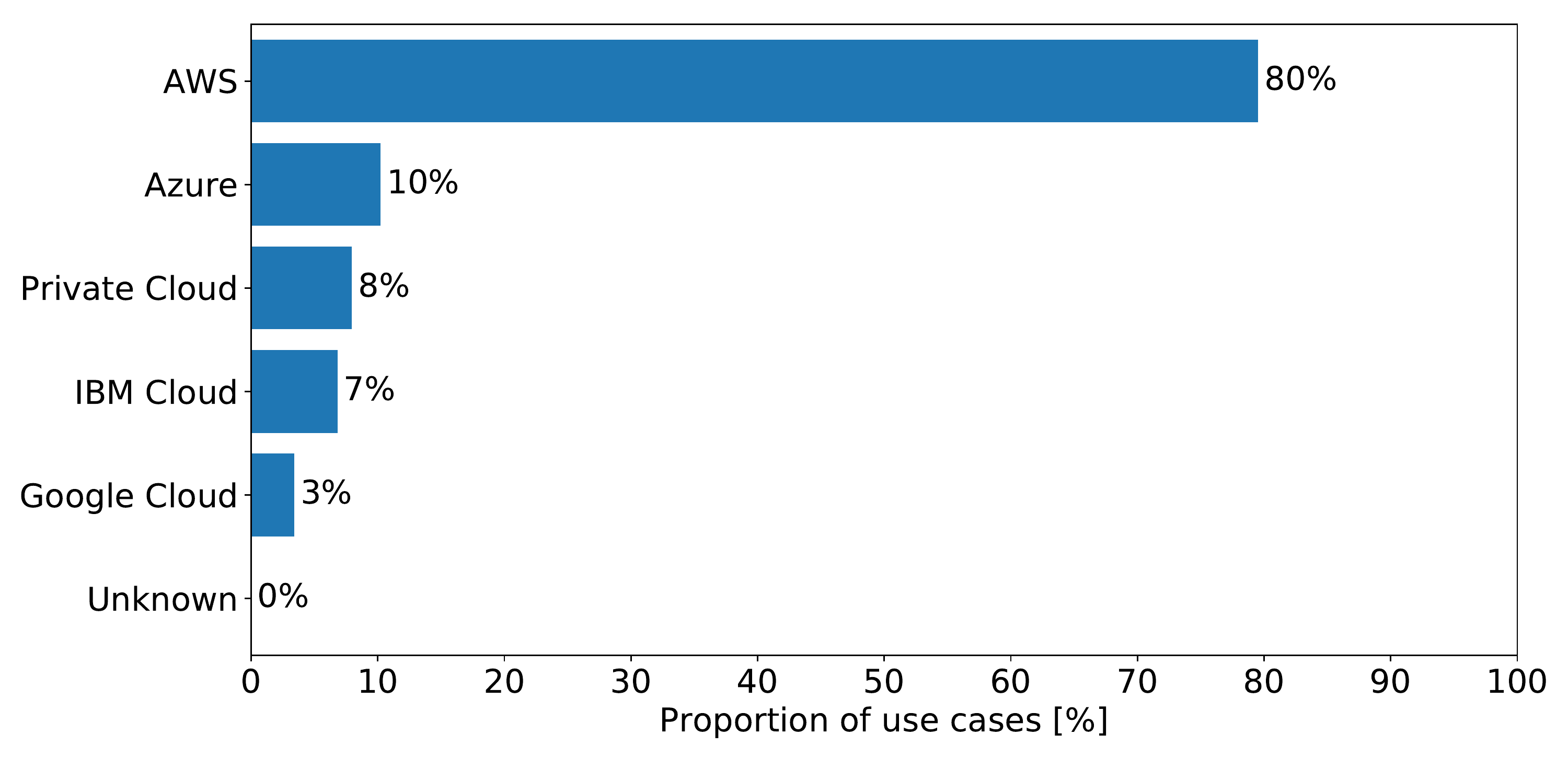}
    \caption{Distribution of deployment platform among the surveyed use cases. Some use cases support multiple deployment platforms.}
    \label{fig:platform}
\end{figure}

\para{Results}
Among the use cases we surveyed, AWS is the clear choice-leader, with 80\% of the use cases choosing AWS as their deployment platform. The other cloud vendors are far behind, with Azure at 10\%, IBM at 7\% and Google Cloud with 3\%. 8\% of the use cases use a private cloud, with the majority of them being scientific use cases. A total of five use cases can be deployed across multiple cloud platforms.\\

\para{Discussion}
That AWS is by far the most popular choice for serverless deployment among the cases we surveyed can probably be attributed to AWS having a two year head-start with offering this technology as a commercial service.
A consequence of the earlier head-start is that 
there was more time to develop and report about serverless applications using AWS serverless technology. Additionally, AWS has the largest market share when it comes to general cloud computing~\cite{gartner}, which gives it a larger existing user base that can move applications to serverless.\\

The very low adoption of private clouds outside of the scientific workflows is in strong contrast to the large number of open-source Function-as-a-Service frameworks that have been developed. A large appeal of the serverless application model is the reduction of operational concerns, so we hypothesize that the increase in operational concerns that comes with maintaining a fleet of servers and an open-source Function-as-a-Service frameworks is deterring the adoption of these frameworks. Additionally, most serverless applications make use of many managed services (storage, databases, messaging, logging, streaming, etc.) which are not available directly in a private cloud environment.\\

It is interesting that five of the use cases we studied can be deployed across multiple cloud platforms. This goes in contrast to the commonly reported vendor lock-in of serverless computing~\cite{adzic2017serverless, eivy2017wary}. However, upon closer inspection, three of these use cases are computation frameworks that provide additional layers of abstraction on top of commercial Function-as-a-Service and another one is a monitoring framework that utilizes serverless technologies. Therefore, we conclude that while there are some frameworks that can operate across multiple cloud platforms, most serverless applications can only be deployed on the cloud platform they were initially developed for. This provides further evidence that vendor lock-in exists for serverless applications.

\subsubsection{Application Type} 
\label{subsec:application_type}
\para{Description}
We categorized each use case according to their type of serverless application. 
The motivation behind this was to explore for what kind of tasks a serverless approach is typically employed---whether there are certain application types that are dominant or application types that are notably missing in current use cases.
To evaluate this usage aspect, we added the \textit{Application Type} metric in which we provided options of typical application types to group the use cases in:

\begin{enumerate}
    \item \emph{Operations \& Monitoring:} Consists of the use cases that deploy serverless application to assist in operating software systems. Examples of such applications include automation of the test or deployment pipelines, failure mitigation or remediation, or controlling the state of running systems. This label superseeds all other labels.
    \item \emph{Stream/async Processing:} Groups the serverless applications that perform an asynchronous task, which also includes any processing of events from an event bus or stream.
    \item \emph{Batch Task:} This is a special case of the \texttt{stream/async processing} category, which encompasses tasks that are executed in large batches. This label superseeds the \texttt{stream/async processing} label.
    \item \emph{API:} Contains use cases that employ serverless to implement an API, such as a REST API or a GraphQL API. The exact nature of this API is not relevant here, rather that it is called synchronously, so the caller is waiting for a response.
    \item \emph{Unknown:} Denotes the use cases that did not provide enough information about what type of serverless applications were employed.
\end{enumerate}

\begin{figure}[ht]
    \centering
    \includegraphics[width=0.8\linewidth]{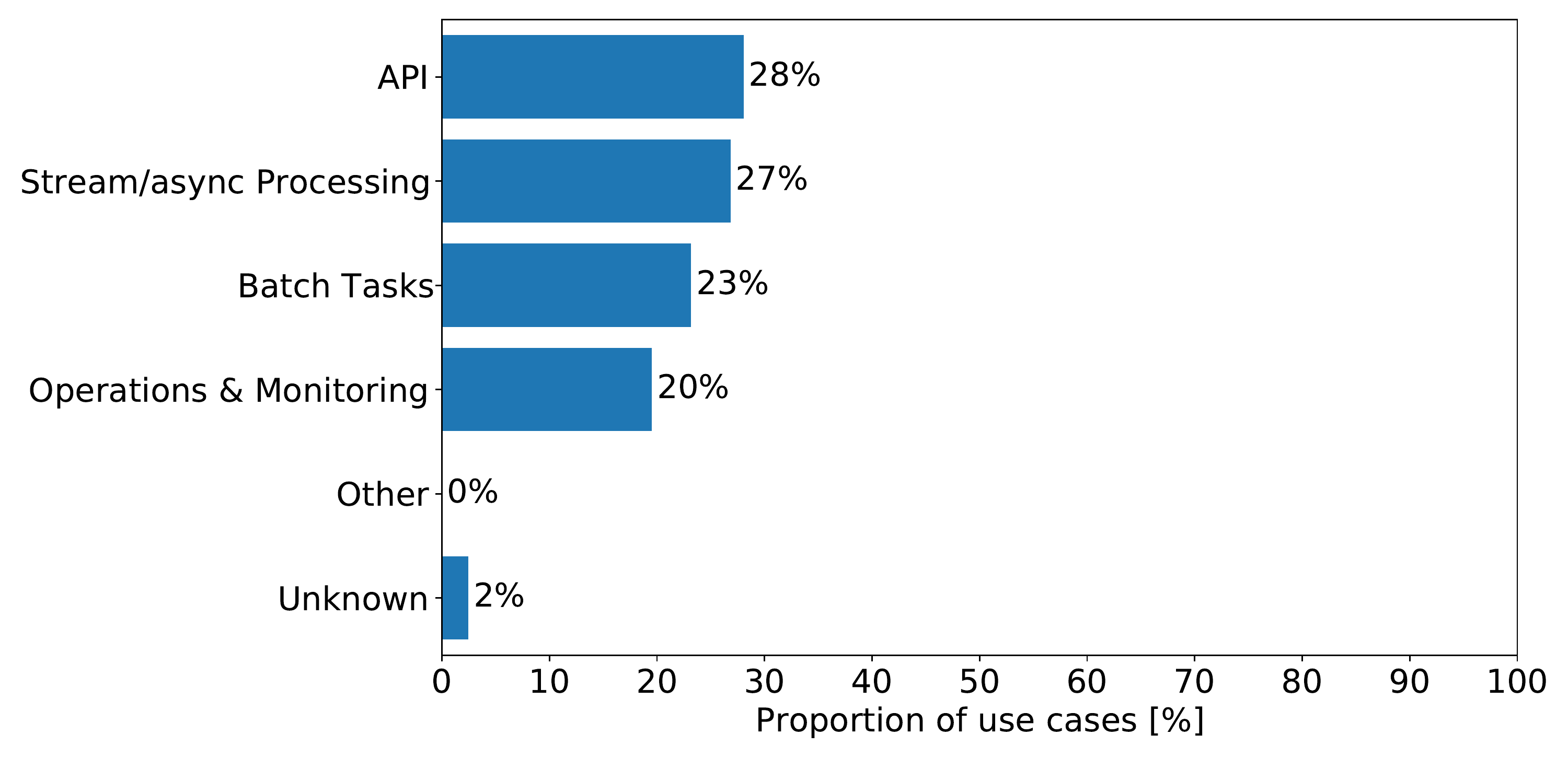}
    \caption{Application type distribution among the surveyed use cases.}
    \label{fig:application_type}
\end{figure}
\para{Results}
Figure~\ref{fig:application_type} provides an overview of the collected results for the \textit{application type} metric. We find that serverless is commonly used to implement \texttt{APIs} (28\%), \texttt{stream/async processing} (27\%), \texttt{batch tasks} (23\%), and \texttt{operations/monitoring tasks} (20\%). We were able to determine this characteristic for all but 2\% of the survey applications.

\para{Discussion}
Serverless is commonly recommended for operations tasks, which also shows in our results as 20\% of the use cases implement operations and monitoring applications. However, we also find large shares of APIs, asynchronous processing, and batch tasks. This shows that serverless is not seen as a niche technology that fits a special use case, but rather as a broadly applicable solution.\\

\subsubsection{In Production}
\label{subsec:production}
\para{Description}
Another characteristic that we analyzed is whether or not a specific use case  was actually deployed in production environments.
Our motivation behind this is to evaluate how reliable the given characteristics actually are, or how representative they are for the real applications running in practice.
Therefore, this metric can be seen as a kind of validation of our results.

For this characteristic, we work with three possible values: 
\begin{itemize}
    \item \emph{Yes}: There is clear evidence or statements claiming that the specific use case is already deployed in production.
    \item \emph{No}: There is strong evidence that the specific use case is not deployed in production.
    \item \emph{Unknown}: There can be no information found to support either ``Yes" or ``No".
\end{itemize}
However, in our analysis an ``Unknown" has almost certainly to be seen as a ``No", as if we can not find evidence supporting that a use case was not applied in production, we have to assume that it is not.\\

\begin{figure}[ht]
    \centering
    \includegraphics[width=0.8\linewidth]{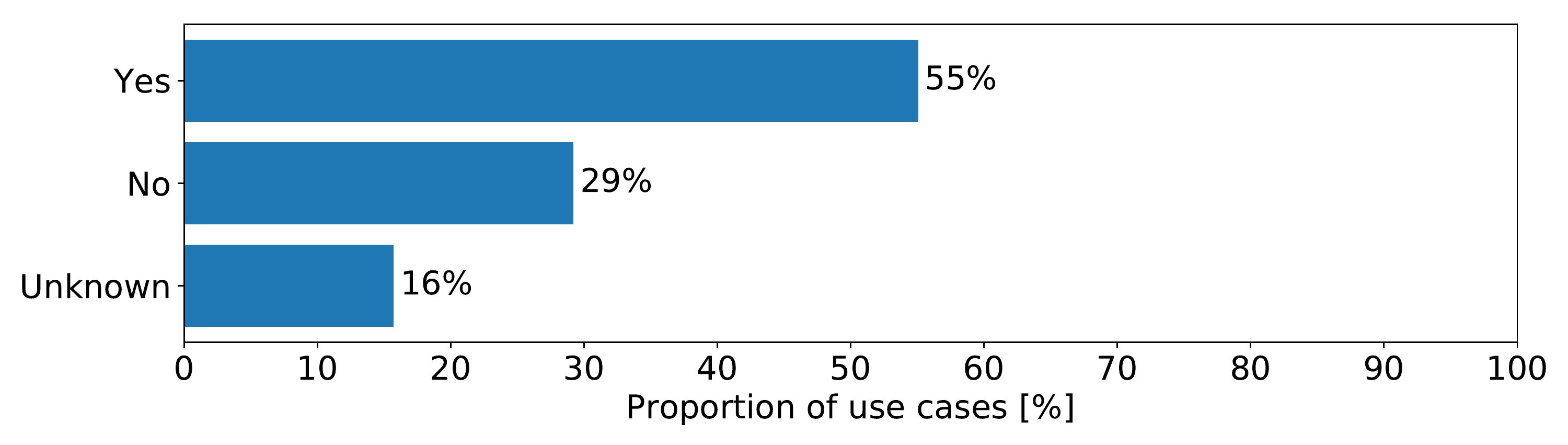}
    \caption{Percentage of the survey use cases that are deployed in production.}
    \label{fig:production}
\end{figure}

\para{Results}
The results of this characteristic are shown in Figure~\ref{fig:production}.
We observe that more that half (55\%) of analyzed use cases are actually designed for and deployed in production.
Roughly a quarter (29\%) is not used in production, and for the remaining 16\% of use cases no clear answer could be found. 
However, as discussed above, we should assume that all ``Unknown'' use cases belong to the ``No'' field, leaving us with a total of 45\% that do not have strong claims supporting their application in production.\\

\para{Discussion}
As more than half of the use cases included in this study are actually used in production, this can be seen as a strong indicator that the results of our analysis are representative and that the developed best practices have been applied in practice.
Furthermore, many of the approaches not actively used in production actually originate from white literature. 
As most of the white literature papers just present prototypical studies and evaluation use cases, their share of non-production use-cases is significantly higher. 
However, this in turn increases the respective share of in-production use cases for grey literature and GitHub Projects.

\subsubsection{Open Source}

\para{Description}
Open source indicates whether the source code of the FaaS function or application is publicly available.
Open source software is a valuable contribution for education, reuse, and testing.
This is exemplified through FaaS providers sharing their own vision for high-level reference architectures\footnote{\url{https://aws.amazon.com/lambda/resources/reference-architectures/}} and fostering cataloging of example applications\footnote{\url{https://aws.amazon.com/serverless/serverlessrepo/}}. \emph{Yes} indicates that a use case is open-source and \emph{No} indicates that no source code is available.
Notice that we barely check the availability of open source artifacts and cannot make any claims about completeness, maintenance levels, or use of appropriate licenses for this characteristic.\\

\begin{figure}[ht]
    \centering
    \includegraphics[width=0.8\linewidth]{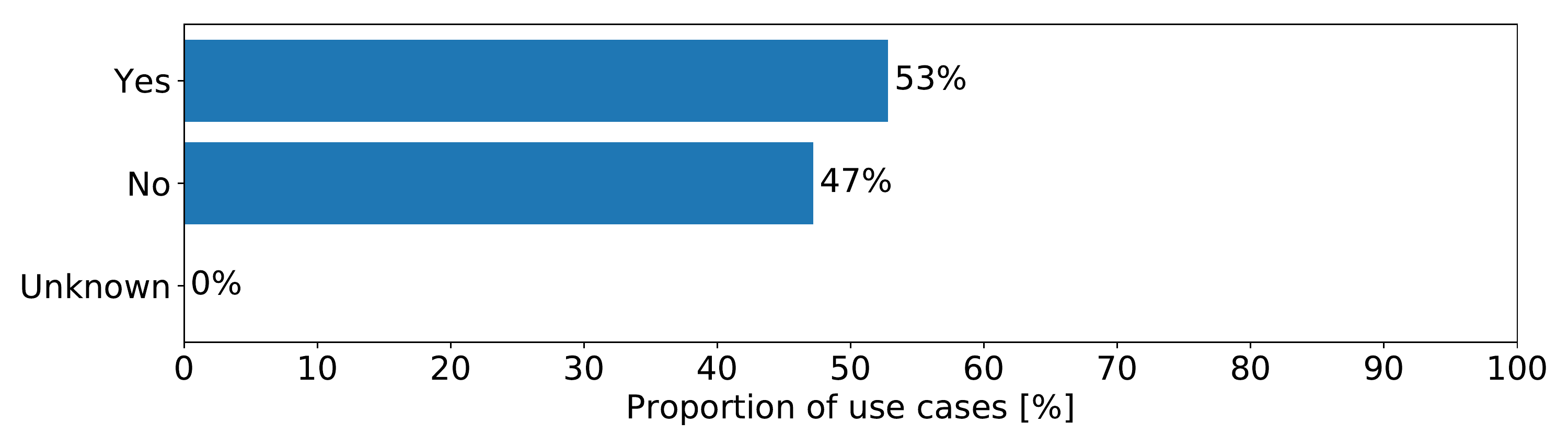}
    \caption{Percentage of the survey use cases that are open-source.}
    \label{fig:open_source}
\end{figure}

\para{Results} Figure~\ref{fig:open_source} shows that 53\% of the use cases are open source and 47\% are closed source. \\

\para{Discussion}
Open source software is typically hosted on GitHub and similarly common for use cases deployed in production.
Interestingly, open source software is comparably widespread among the 49 use cases deployed in production with 49\%.
We expected a clearer tendency of use cases deployed in production to remain closed source, which is probably due to our selection strategy favoring open source use cases.

\subsubsection{Domain} 
\para{Description}
For each use case, we classified their application domain based as one of the following: \emph{IoT}, \emph{entertainment}, \emph{scientific computing}, \emph{WebServices}, \emph{public authority}, \emph{university}, \emph{FinTech}, \emph{cross-domain}, or \emph{other}.
Cross-domain are those use cases that are generic and could be useful across more than one domain; for example, a generic image identification service which could be used in IoT, scientific computing, WebServices, public authority, and university domains.
\\

\begin{figure}[ht]
    \centering
    \includegraphics[width=0.8\linewidth]{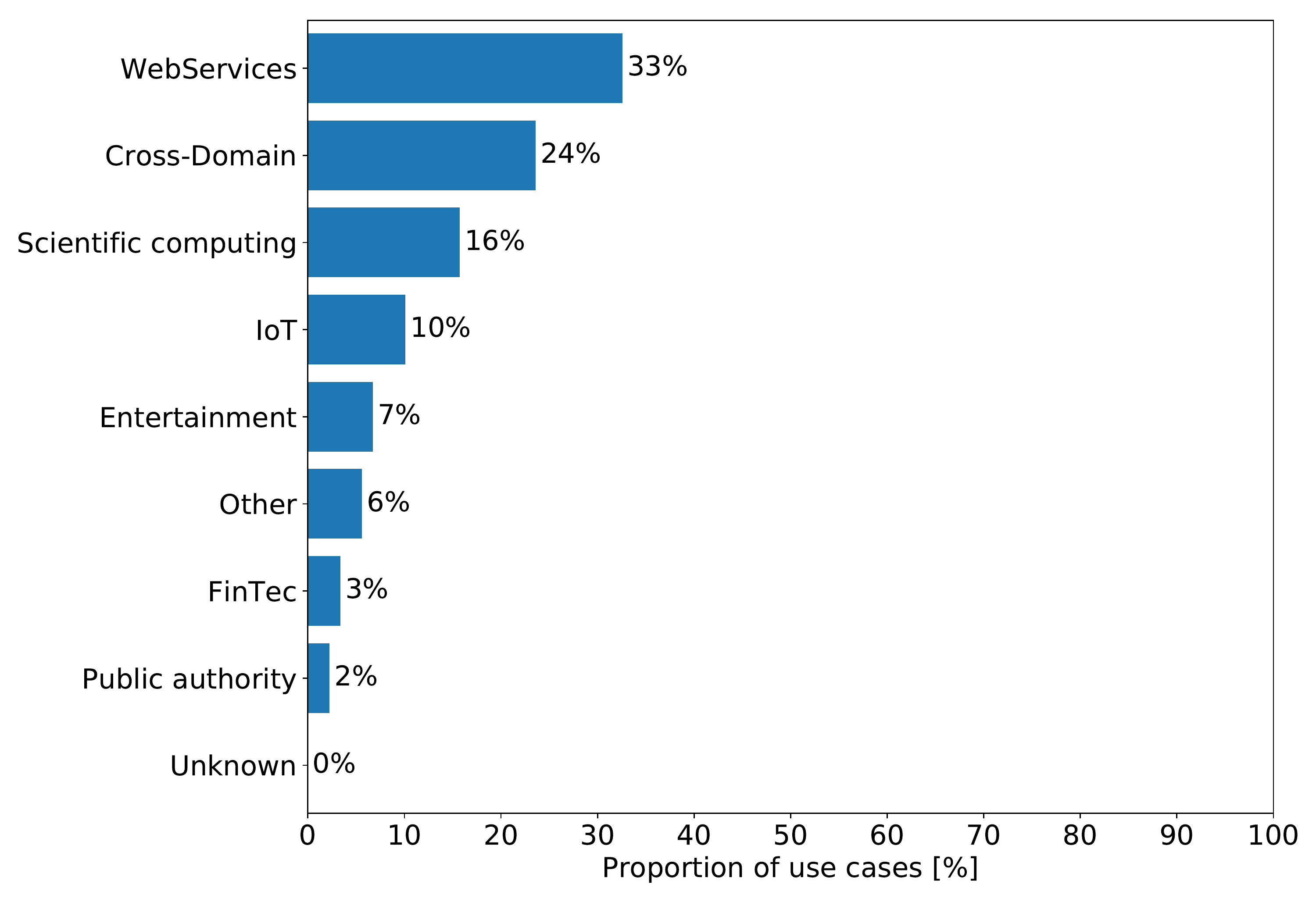}
    \caption{Distribution of the domain of the surveyed use cases.}
    \label{fig:domain}
\end{figure}

\para{Results}
Unsurprisingly, WebServices is the most common application domain in our survey (33\%, see Figure~\ref{fig:domain}).
This is followed by cross-domain use cases (24\%), which we mostly found via our GitHub search.
The other groups with high representation in our review were scientific computing (16\%) and IoT (10\%).
The significant presence of scientific computing cases is a result of our conscientious efforts in including scientific computing use cases in our survey.
\\

\para{Discussion}
We find that there is a wide variety of application domains represented in our survey.
This is a strength of our study, as others can use our insights to make decisions about the design and implementation of serverless frameworks that are applicable to a broad variety of domains.

\subsection{Workload Characteristics}
\label{sec:results:workload}

This section characterizes the nature of workloads imposed on serverless applications through human users or technical invokers.
In the following, we discuss execution patterns, burstiness, trigger types, and common data volumes of our reviewed use cases.

\subsubsection{Execution Pattern}

\para{Description}
Functions or workflows of functions can be triggered \emph{on-demand} as a direct result of a user interacting with the application, or they can be \emph{scheduled} to be run at specific times.
For the on-demand workflows, we further classify them as regular \emph{on-demand} or \emph{high-volume on-demand}.\\

\begin{figure}[ht]
    \centering
    \includegraphics[width=0.8\linewidth]{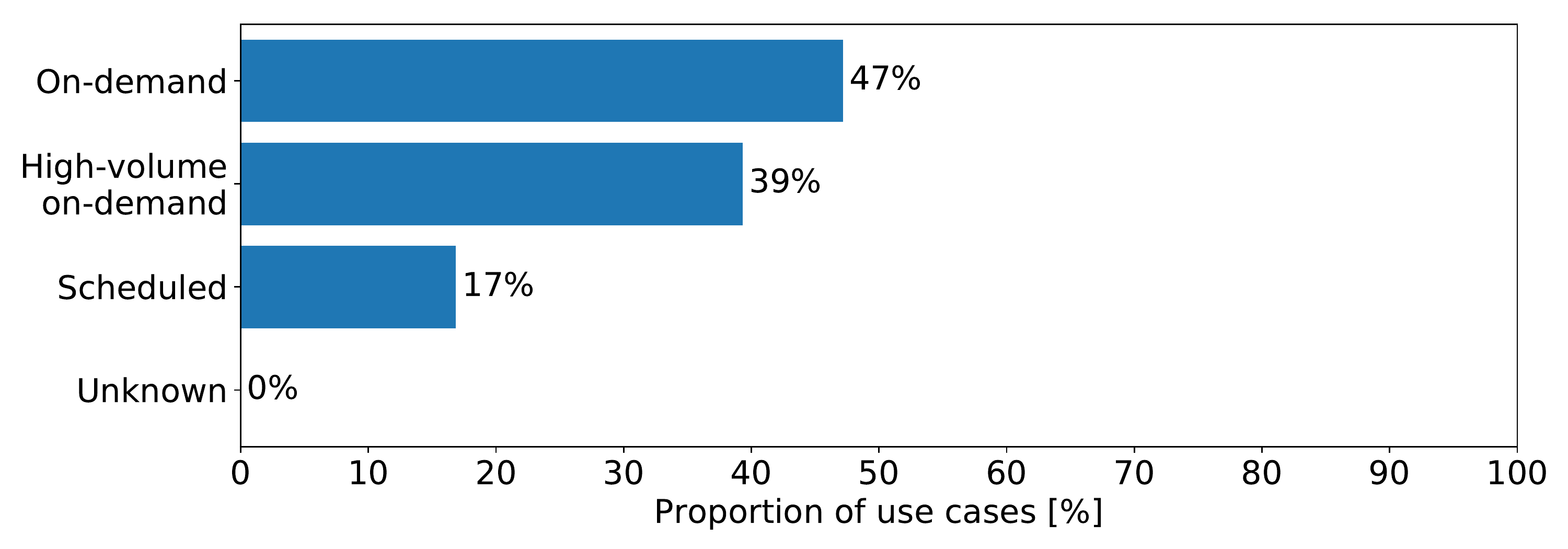}
    \caption{Execution pattern distribution among the surveyed use cases. Some use cases are executed both on a schedule and on-demand.}
    \label{fig:execution_pattern}
\end{figure}

\para{Results}
Most (workflows of) functions are triggered on-demand, with scheduled triggers being used in only 17\% of the use cases we analyzed (see Figure~\ref{fig:execution_pattern}).
Out of the on-demand execution patterns, close to half are high-volume, business-critical invocations. \\

\para{Discussion}
The high prevalence of high-volume on-demand triggers calls for special study in minimizing function start-up times, and auto-scaling mechanisms.
In addition, half of the workflows that are scheduled fall into the application type category \emph{operations} (see \cref{subsec:application_type}), highlighting how the serverless model has been adopted---in many cases---to automate operations, software management, and DevOps pipelines.
\\

\subsubsection{Burstiness}
\para{Description}
The workload of a function can be either bursty or non-bursty.
A bursty workload follows a workload pattern that includes certain sudden and unexpected load spikes, or alternatively a significant amount of sustained noise and variation in intensity.
We classify a use case as bursty (i.e., \emph{yes} for burstiness) if its workload typically includes or can include burst patterns in some situations and as non-bursty (i.e, \emph{no}) if the workload is almost guaranteed to never receive burst patterns (e.g., if all executions are scheduled and known in advance).
If the burstiness of a use case is \emph{unknown}, then the use case was either under-specified, or can be both bursty or non-bursty depending on the specific area of application.
Note that in any scenario that involves a set of human users, we consider the workload pattern to be bursty, as user behavior can almost never the scheduled or reliably controlled, leading to a possibly bursty behavior.\\

\begin{figure}[ht]
    \centering
    \includegraphics[width=0.8\linewidth]{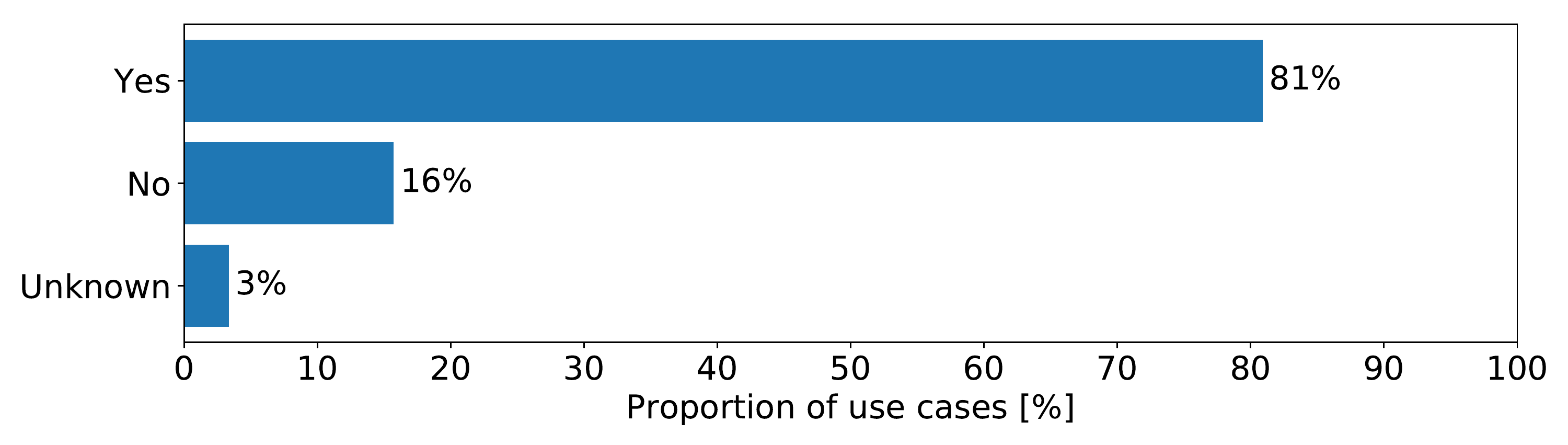}
    \caption{Percentage of the survey use cases that have a bursty workload.}
    \label{fig:bursty}
\end{figure}

\para{Results}
Figure~\ref{fig:bursty} depicts that a large majority (81\%) of the analyzed workload patterns are classified as bursty.
Additionally, a 16\% share have a clear non-bursty workload pattern, while a small minority of 3\% could not be attributed to be either bursty or non-bursty.\\

\para{Discussion}
As one of the strengths of serverless computing is its seamless and almost infinite scalability, together with the general ease of operations it comes as no surprise that most of the use cases indeed experience bursty workload patterns.
Early adopters that want to test out the new emerging paradigm are more likely to choose a use case that is optimized for the serverless offering. 
At the same time, engineers that have been struggling with bursty workloads and face regular performance issues are also more likely to migrate or adopt their application towards the a serverless clouds than applications that run smoothly.

An interesting comparison here would be to compare the share of bursty workloads executed on serverless platforms versus the share of bursty workloads of conventional applications.
However, we can still conclude that a large majority of use cases designed for or applied to serverless platforms are experiencing bursty workloads and hence make use of the seamless elasticity that these services offer.

\subsubsection{Trigger Types}
\label{subsec:trigger_types}

\para{Description}
Trigger types refer to alternative ways of invoking a FaaS function and are closely related to external services (see \Cref{subsec:external_services}).
An \emph{HTTP request} can trigger a FaaS function, which then processes the request and generates an HTTP response.
This HTTP routing is often implemented through API gateways.
A \emph{cloud event} describes a state change happening in a connected cloud service, such as a file upload to cloud storage or a modified value in a cloud database.
Such cloud events can be configured to trigger new function executions.
A \emph{scheduled} trigger invokes a FaaS function at a defined and potentially recurring time.
The category of \emph{manually} triggered functions refers to human-initiated executions typically executed on-demand.
Notice that some use cases combine multiple trigger types and thus the sum of proportions exceeds 100\%. \\

\begin{figure}[ht]
    \centering
    \includegraphics[width=0.8\linewidth]{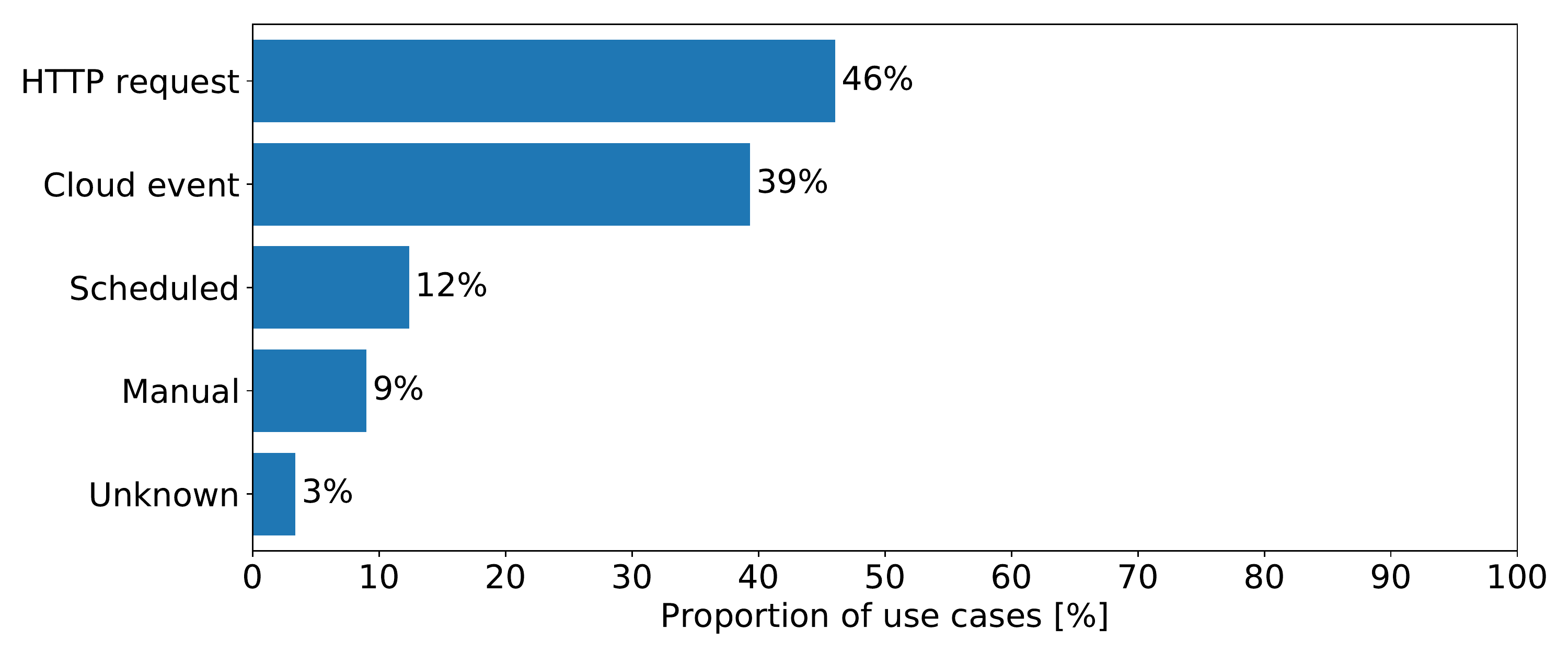}
    \caption{Trigger type distribution among the surveyed use cases. Some use cases have multiple trigger types. }
    \label{fig:trigger}
\end{figure}

\para{Results}
Figure~\ref{fig:trigger} reveals that the most common trigger types are HTTP request (46\%) and cloud event (39\%).
Far less common are scheduled (12\%) and manual (9\%) execution triggers.
We were unable to derive the trigger type for 3\% of the use cases from their insufficient descriptions.\\

\para{Discussion}
We compare our results to the trigger types reported for the production workload of Microsoft Azure functions~\cite{shahrad2020serverless}.
Scheduled triggers and HTTP triggers are both 16-20\% more common among Azure FaaS applications in comparison to our analyzed use cases focusing on AWS (85\%).
The results for the remaining categories very closely (<=2\%) match (after mapping some cumulative categories).
We conclude that the order of categories is in line with current production workloads reported for Azure but note that some values might be higher in practice.
Such an underestimation is plausible given that we derive our results from potentially incomplete sources.
However, the explicit grouping of functions into applications in the Azure FaaS implementation possibly leads to different function groupings compared to our AWS-focused use cases.

\subsubsection{Data Volume}

\para{Description}
The data volume defines what load will be on network and storage devices. 
The motivation here is to analyze whether there are any clusterings of data usages or certain patterns that are generally avoided.
We categorized the different use cases into five different categories: Volumes of \emph{less than 1 MB} per execution, \emph{less than 10 MB}, \emph{less than 100 MB}, \emph{less than 1 GB}, and \emph{more than 1 GB}.
Additionally, there is also the \emph{unknown} category, if data volume could not be assessed.
Note that the data volume refers to executions of the entire workflow.
Furthermore, as exact numbers were seldom found in the sources, this categorization is often based on the estimate of our reviewers.\\  

\begin{figure}[ht]
    \centering
    \includegraphics[width=0.8\linewidth]{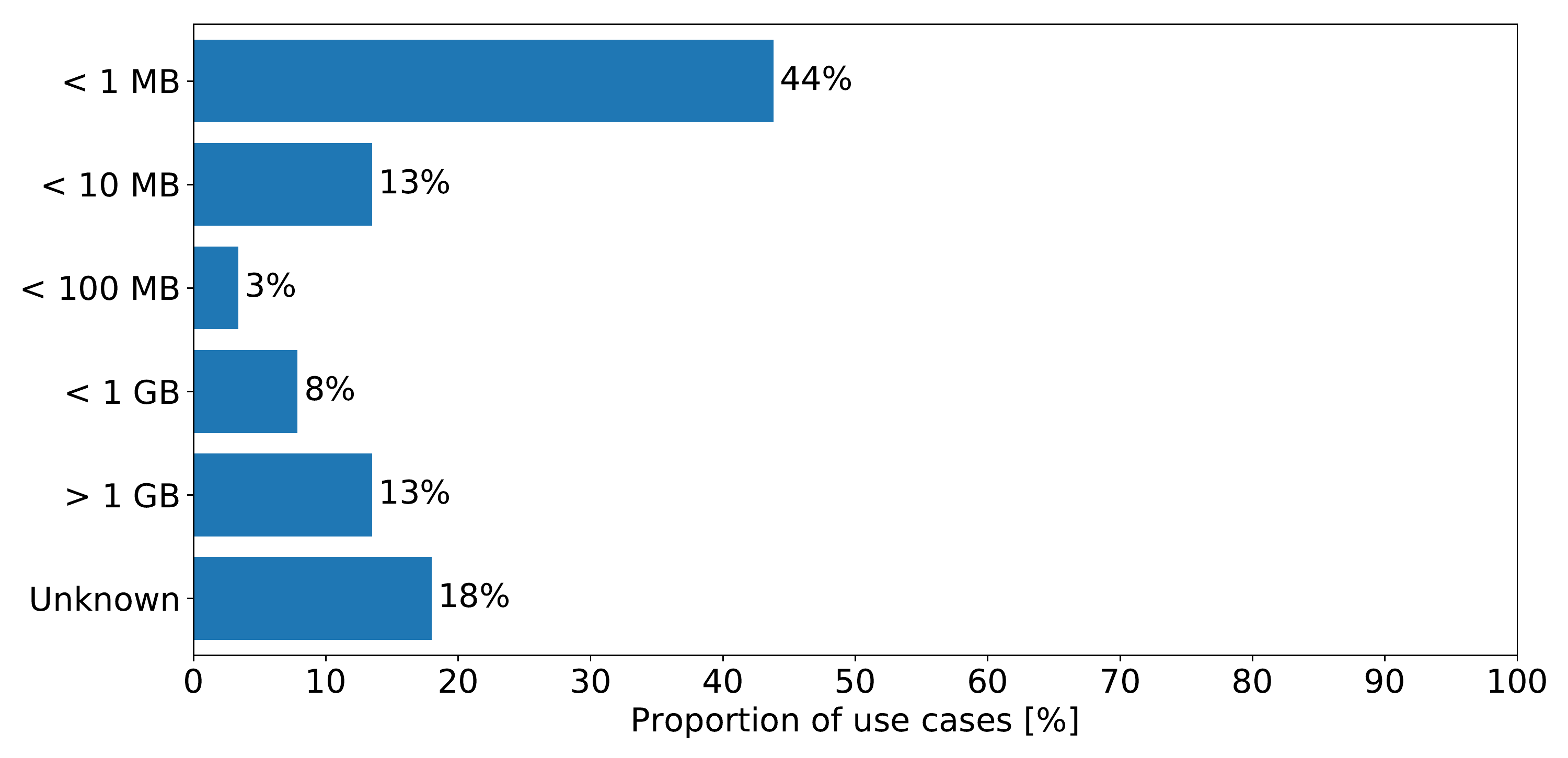}
    \caption{Data volume distribution among the surveyed use cases.}
    \label{fig:data_volume}
\end{figure}

\para{Results}
Figure~\ref{fig:data_volume} depicts the distribution of use cases among the different classes. 
Almost half of the use cases (44\%) fall in the smallest category of data volumes of less than 1 MB. 
The second categorization transmitting more than 1 MB of data, but less than 10 MB, also make the second largest group with a fraction of 13\%.
Even less use cases (3\%) consider a data volume between 10 and 100 MB. 
However, the following group between 100 MB and 1 GB increases in popularity, and finally the share of use cases transmitting 1 GB or more to the serverless platform increases to be the second-largest group (13\%).
Additionally, 18\% of use-cases could not have a specific data volume assigned and therefore do not count into any of the enumerated groups.\\

\para{Discussion}
Generally, the different data volumes are relatively distributed and do not cast a clear picture.
There is definitely a use case for any data volume characteristic. Therefore platforms should not strive to optimize themselves towards any specific limitation here.
That said, most of the use cases transmit less than 1 MB of data per workflow execution. 
Note that this group also includes all use cases that might not send any data at all. 
Therefore, the large majority of serverless use cases that we surveyed does not work with big amounts of data. 
However, there is also the exact opposite group of use cases working with vast vast amounts of data of 1 GB and more per workflow execution.

\subsection{Application Characteristics}
\label{sec:results:application}
This section characterizes the how the applications use cloud functions.
In the following, we analyze the applications regarding: the number of distinct functions in them, the function run times, the resource bounds of the functions, the programming languages used to implement the functions, the upgrade frequency of the cloud functions, and their interactions with external cloud services.

\subsubsection{Number of Distinct Functions}
\para{Description}
The business logic of serverless applications is contained within serverless functions and connects to a variety of managed cloud services. Similarly to microservices, the appropriate granularity of serverless functions is a controversial topic. Opinions range from wrapping each programming function as a serverless function, each API endpoint as a serverless function to full microservices as a serverless function. In this characteristic we investigate the number of distinct functions within the use cases. As this characteristics targets the development perspective, we count a function that is executed multiple times within an application as a single function. For some use cases, the only information available is along the lines of ''more than X functions", which we count as X for this analysis.\\

\begin{figure}[ht]
    \centering
    \includegraphics[width=0.8\linewidth]{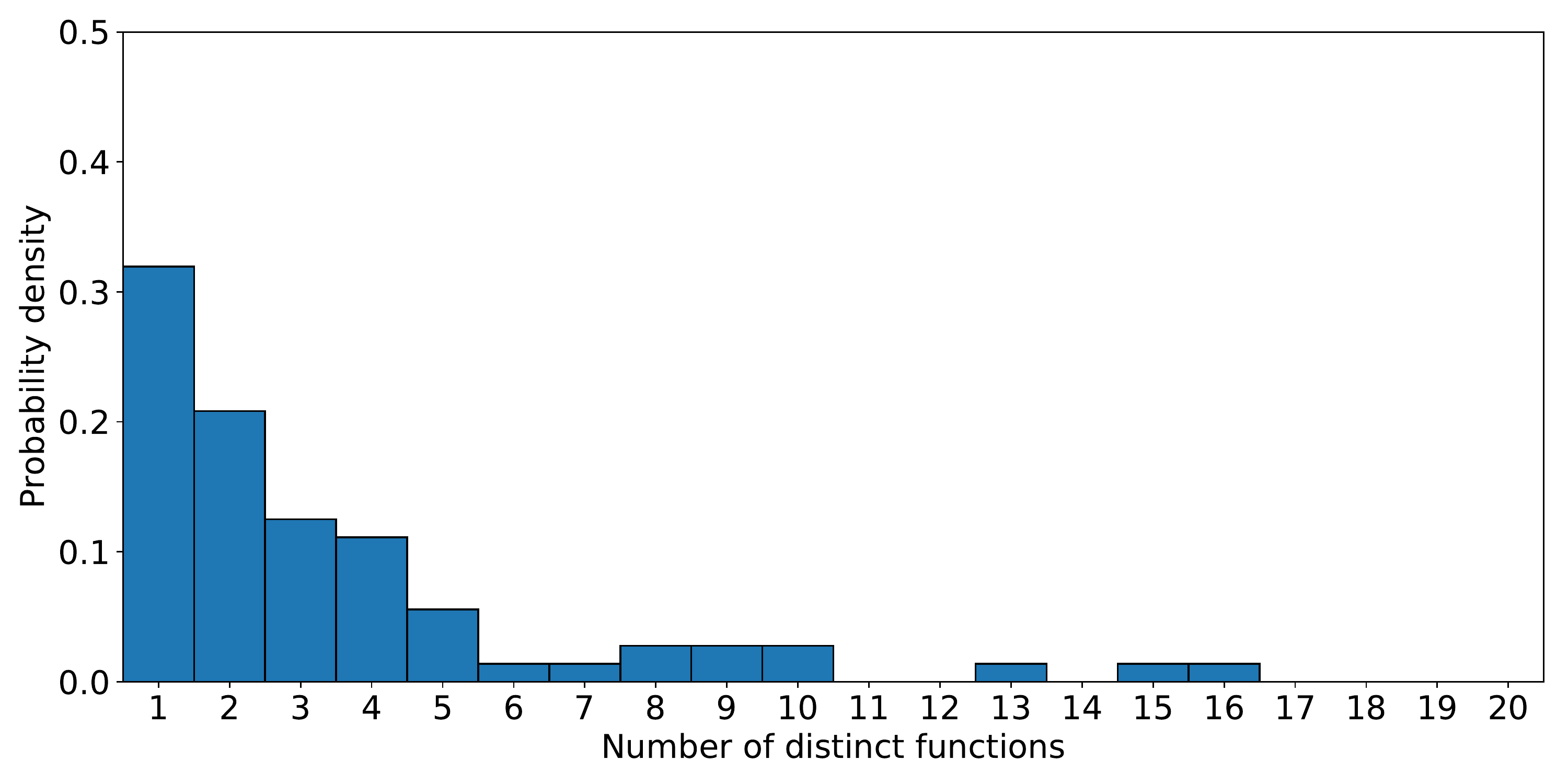}
    \caption{Histogram of the number of functions per use case, single outlier at 170 not shown.}
    \label{fig:number_of_functions}
\end{figure}

\para{Results} About a third (32\%) of the analyzed use cases consist of only a single function, as shown in %Figure~
\autoref{fig:number_of_functions}.
Further, about one-fifth (21\%) consist of two functions, a tenth (12\%) of three functions, another tenth (11\%) of four functions, and 5\% of five functions. 
Larger sizes are very rare and without causing a mode in the empirical distribution: there exists in our analysis only one use case for each of the sizes 6, 7, 13, 15, 16, and, notably, 170; there exist only two use-cases for each of the sizes 8, 9, and 10. The use case with more than 170 functions is the back-end for the mobile app of a now defunct start-up. 
Overall, 82\% of all use cases consist of five functions or less. 
Furthermore, 93\% of the use cases that consists of ten functions or less. \\

\para{Discussion}
Our results determine that serverless application use a low number of serverless functions, with 82\% of all use cases consisting of five or less functions and 93\% of the use cases consisting of less than ten functions. There are two potential reasons for this. First, the serverless application models reduces the amount of code developers have to write, as it allows them to focus on business logic while all other concerns are taken care of by the cloud provider and managed cloud services. Secondly, this seems to indicate that developers are currently choosing a rather large granularity for the size of serverless functions. However, determining the optimal granularity for serverless functions is still an open research challenge.

\subsubsection{Function Runtime}
\para{Description}
The run time of the cloud functions may have important impact on optimization choices of the serverless frameworks running these functions. We classified the run time of the functions in the use cases as: \emph{short} (order of milliseconds or seconds) and \emph{long} (order of minutes).\\

\begin{figure}[ht]
    \centering
    \includegraphics[width=0.8\linewidth]{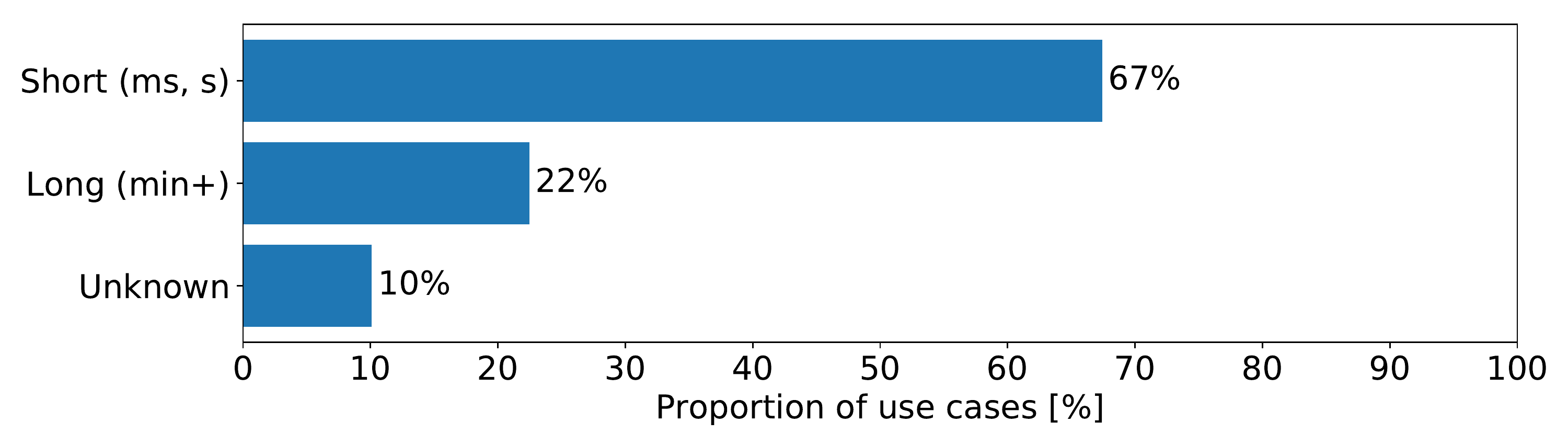}
    \caption{Function runtime distribution among the surveyed use cases.}
    \label{fig:function_runtime}
\end{figure}

\para{Results}
The majority of the functions in our survey are short (ms, s), 67\%; only 22\% of them have a run time in the order of minutes (see Figure~\ref{fig:function_runtime}). We could not assess this characteristic in 10\% of the use cases we studied. 
All but one of the long-running functions is triggered on demand (as opposed to scheduled), with half of them falling into the scientific computing domain.
The long-running functions that did not fall within the scientific computing domain, are mostly operations or side-tasks, not business critical.
Finally, these long functions are not high-volume on demand APIs (only one was classified as such).\\

\para{Discussion}
The overhead associated with running a function is larger, in proportion, for the case of functions with short run times.
This supports the large number of efforts concentrated in reducing this overhead. 
A limitation of our results is that, as the platforms impose a run time limit in the order of minutes, there may be a bias towards short running functions that would not exist if there were no time limits to the function run times. \\

\subsubsection{Resource Bounds}
\para{Description}
We wanted to know if the functions' run time were limited by \emph{I/O}, \emph{CPU}, both (\emph{hybrid}), the \emph{network}, or an \emph{external service} (e.g., cloud database).
Information on the workload mix can be useful for studies regarding the scheduling of functions and the routing of execution requests. \\

\begin{figure}[ht]
    \centering
    \includegraphics[width=0.8\linewidth]{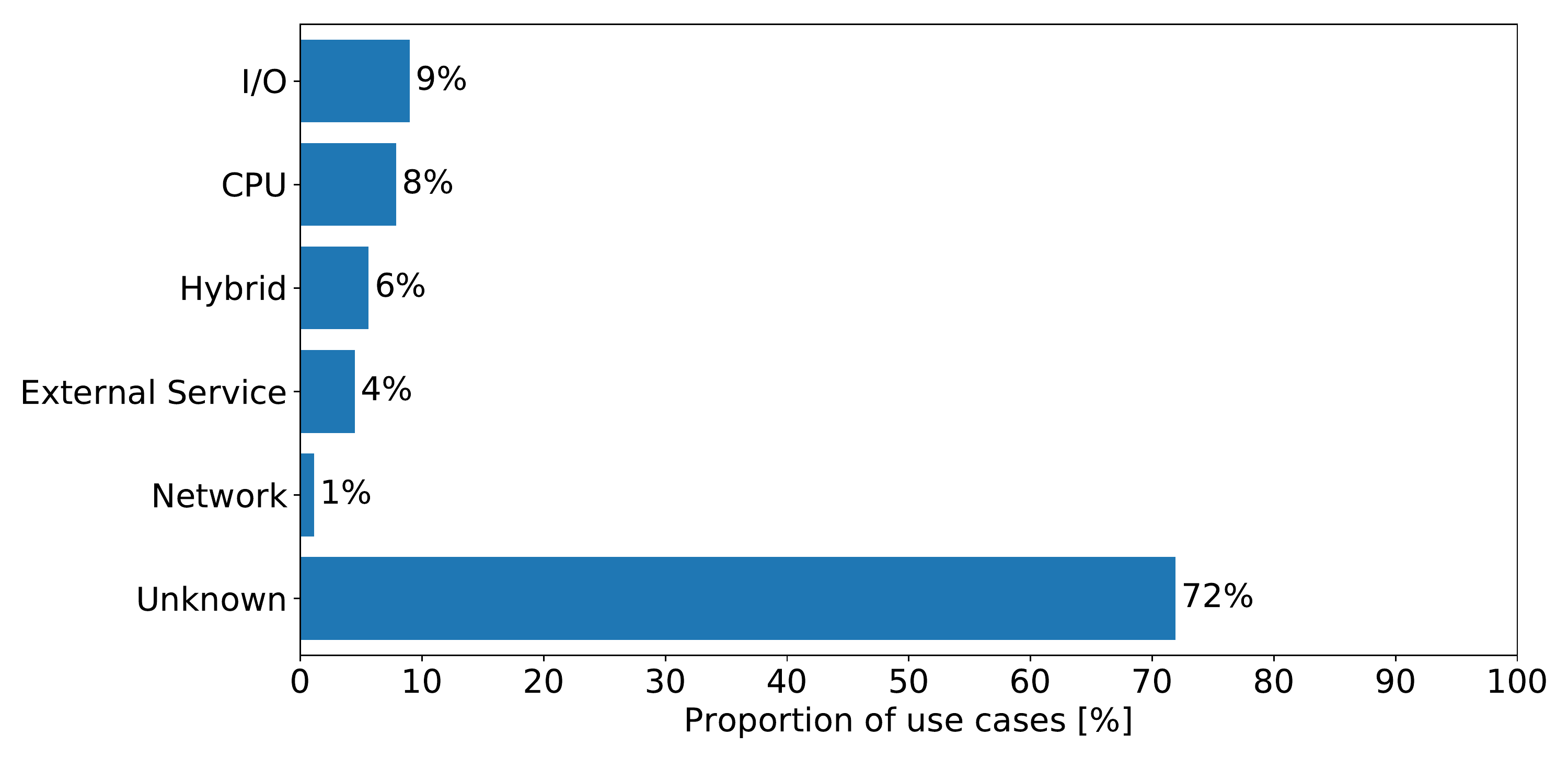}
    \caption{Distribution of the resource bounds among the surveyed use cases.}
    \label{fig:resource_bounds}
\end{figure}

\para{Results}
Most use cases did not explicitly state this information (\emph{unknown} is 72\%, see Figure~\ref{fig:resource_bounds}). For the use cases that did report this information, the I/O-bound functions and CPU-bound functions are equally represented in our survey (9\% I/O, 8\% CPU, 6\% hybrid, 4\% external service and 1\% network). \\

\para{Discussion}
The percentage of use cases reporting this information is too small for us to derive any statistically significant analysis of the results. \\

\subsubsection{Programming Languages}
\para{Description}
This characteristic refers to the main programming languages used to write code for FaaS functions in a given application.
FaaS providers typically offer a set of officially supported runtimes (e.g., Node.js for JavaScript).
These execution environments of FaaS functions determine the operating system and pre-installed software libraries.
Some providers support further languages through custom runtimes, often in the form of Docker images following a documented interface.
Notice that the programming language might differ from the technical function runtime as so called \emph{shims} can be used to invoke a target language through a wrapper runtime (e.g., invoking C++ through Node.js via system calls). \\

\para{Results}
JavaScript (32\%) is the most common programming language for FaaS functions tied with Python (32\%).
Less common languages include Java (9\%), C and C++ (8\%), C\# (6\%), Go (3\%), and Ruby (1\%).
We were unable to determine the language for 25\% of the use cases due to lacking technical descriptions. \\

\begin{figure}[ht]
    \centering
    \includegraphics[width=0.8\linewidth]{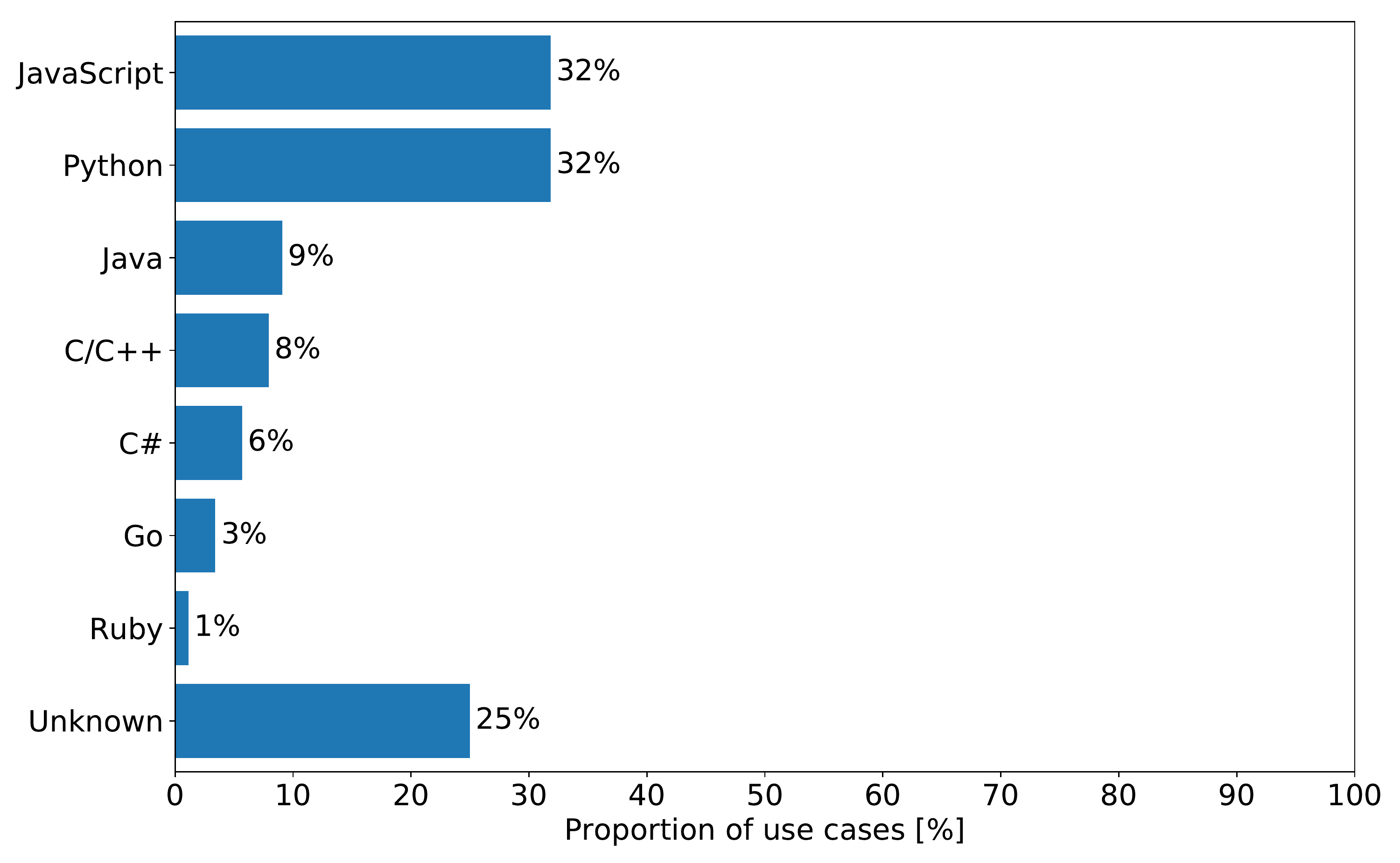}
    \caption{Programming language distribution among the surveyed use cases. Some use cases use multiple programming languages.}
    \label{fig:languages}
\end{figure}

\para{Discussion}
The ranking of languages in our results follows a general trend also observed in other surveys. 
A study on FaaS industrial practices (N=161) \cite{leitner2019MixedMethod} and initial results of the latest Serverless Community survey (N=109)\footnote{Question 25 in \url{https://www.nuweba.com/blog/serverless-community-survey-2020-results}} indicate that JavaScript is 20\% more popular than Python on used languages in FaaS applications.
However, they largely follow the same ranking but suggest higher popularity of Java over C\#.
Our results are plausible and confirm that JavaScript and Python are the most widely supported programming languages in FaaS.
Further, the remaining languages (i.e., Java, C, etc.) all belong to the world's most popular languages according to the TIOBE index\footnote{\url{https://www.tiobe.com/tiobe-index/}}, although they are often not the primary choice for the new FaaS paradigm.

\subsubsection{Function Upgrade Frequency} 
\para{Description}
How frequently the code of the functions is updated has implications to software engineering and to the mechanisms used by the framework to upgrade the code in the functions that are already deployed.
We used two classification levels for this property: \emph{rarely} and \emph{often}; \emph{unknown} indicates that this information cannot be obtained from the use case. \\

\begin{figure}[ht]
    \centering
    \includegraphics[width=0.8\linewidth]{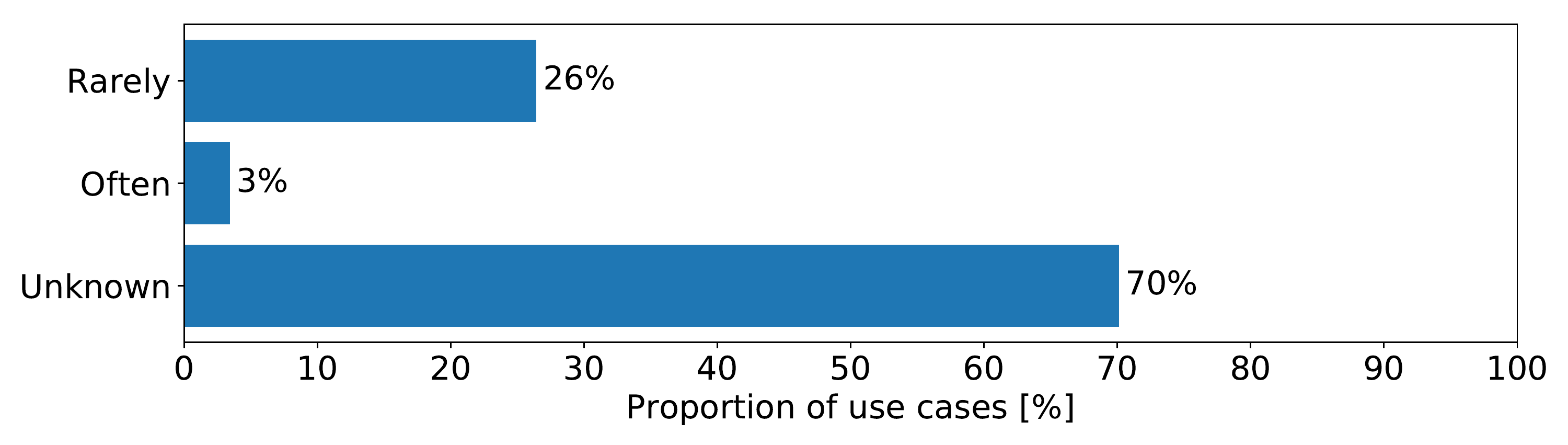}
    \caption{Function upgrade frequency distribution among the surveyed use cases.}
    \label{fig:upgrade}
\end{figure}

\para{Results}
Most use cases did not explicitly state this information (unknown is 70\%, see Figure~\ref{fig:upgrade}). For the use cases that did report this information, the functions are updated rarely (26\% rarely, 3\% often).\\

\para{Discussion}
The percentage of use cases reporting this information is too small for us to derive any statistically significant analysis of the results. \\

\subsubsection{Use of External Services}
\label{subsec:external_services}
\para{Description}
FaaS functions are often integrated into an ecosystem of serverless external services.
Persistency services include cloud storage for blob data (e.g., Amazon S3 for images) and cloud database for structured data storage and querying (e.g., Amazon DynomoDB or Google Cloud SQL).
A cloud API gateway exposes HTTP endpoints and can trigger FaaS functions upon incoming HTTP requests.
Messaging services include cloud pub/sub for durable asynchronous messaging (e.g., Amazon SNS), cloud queue for reliable FIFO-ordered messaging (e.g., Amazon SQS), and cloud streaming for real-time data ingestion and processing (e.g., Amazon Kinesis).
Cloud logging and monitoring refers to applications that explicitly process log data because we implicitly assume some essential logging infrastructure for FaaS functions (e.g., Amazon CloudWatch for AWS Lambda).
Cloud ML covers machine learning services, such as Amazon Rekognition for image or video analysis.
Notice that we abstracted from vendor-specific services to cross-platform terminology (e.g., AWS S3 becomes Cloud Storage). \\
\begin{figure}[ht]
    \centering
    \includegraphics[width=0.8\linewidth]{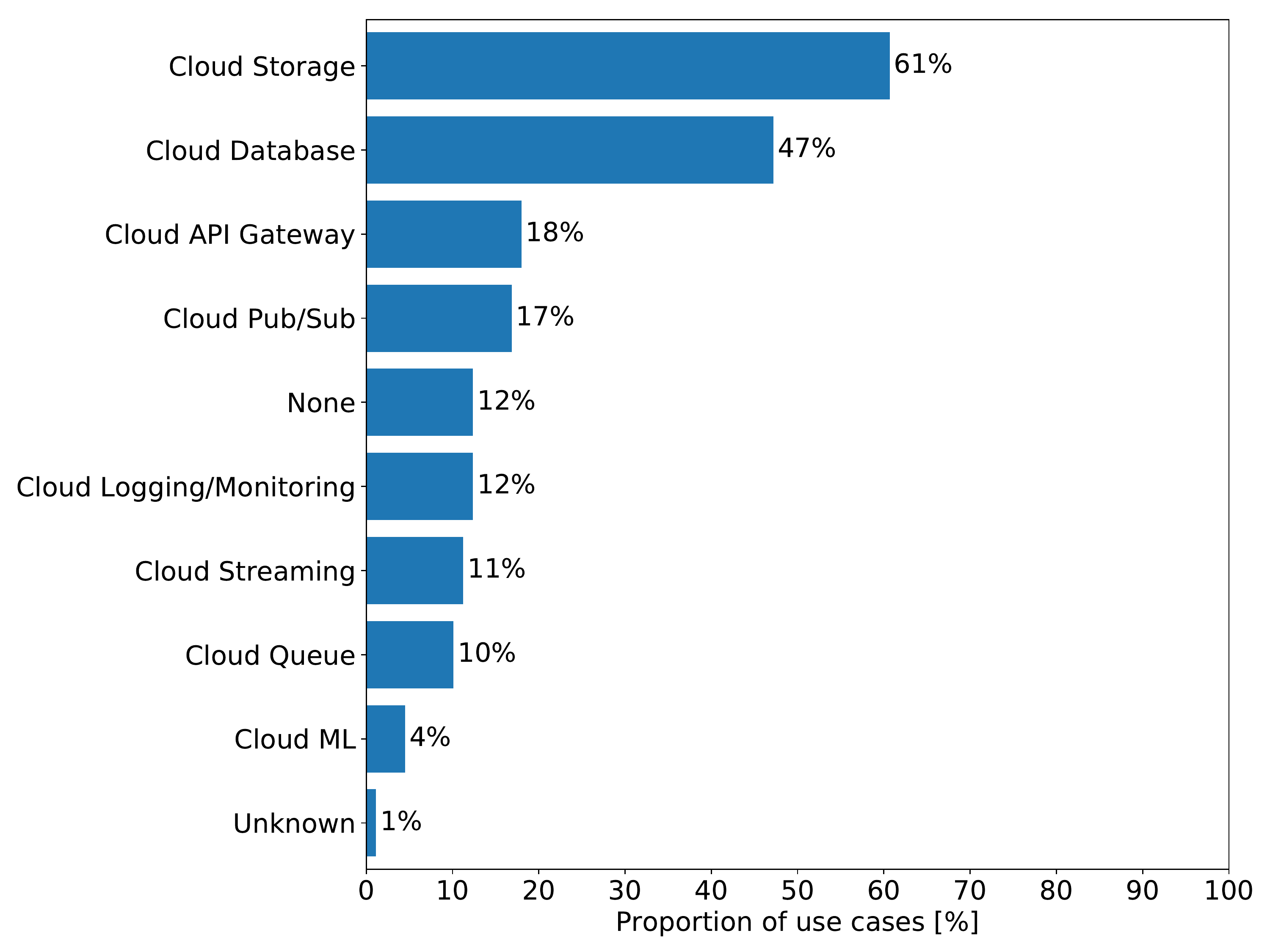}
    \caption{Distribution of used external services among the surveyed use cases. Many use cases use multiple external services.}
    \label{fig:external_services}
\end{figure}

\para{Results}
Figure~\ref{fig:external_services} shows that cloud storage (61\%) and cloud database (47\%) are the most popular external services, followed by the cloud API gateway (18\%) and messaging services (10-17\%).
For 12\% of the use cases, we could not identify any external service integration. \\

\para{Discussion}
Given the ephemeral nature of FaaS functions, it is unsurprising that persistency services are the most popular external services, which is consistent with other survey results~\cite{leitner2019MixedMethod}.
However, the API gateway receives surprisingly little attention, especially when compared to the 46\% of the use cases using HTTP triggers (see \Cref{subsec:trigger_types}).
We suspect the use of an API gateway is often implicitly assumed, thus not explicitly mentioned, and therefore not comprehensively captured here.
Overall, we conclude that currently used external services almost exclusively focus on technical aspects (e.g., storage or messaging) and more specialized services (e.g., cloud ML) are very uncommon among our surveyed use cases.

\subsection{Requirements Characteristics}
\label{sec:results:requirements}
In the following section, we analyze the different requirements and expectations that users have towards the serverless platforms when moving towards those.
We discuss the main motivation drivers, as well as the trade-off between cost and performance, and requirements towards latency and locality of the invocations.

\subsubsection{Motivation}
\label{sec:results:requirements:motivation}
\para{Description}
This characteristic aims at capturing the motivation of the respective engineers and therefore quantifies why they decided to host their application in a serverless environment.
For this, we developed six main motivation fields and grouped each use case into one or multiple of those fields, depending on the motivation the authors gave in the description.
If no conclusive motivation could be found, we put \emph{Unknown}.
The main motivations we found are:
\begin{itemize}
    \item \emph{Cost}: Running the application in a serverless platform significantly reduces operation cost in comparison to traditional cloud hosting.
    \item \emph{NoOps}: Deploying a serverless application has the advantage of saving operation effort.
    \item \emph{Scalability}: The increased scalability of serverless platforms is advantageous for the application.
    \item \emph{Performance}: The performance of the application, i.e., throughputs and response times, is better when running on a serverless platform.
    \item \emph{Simplify Development}: The development cycle as well as the release structure is easier using serverless applications.
    \item \emph{Maintainability}: Deploying an application in a serverless cloud saves maintenance effort.
    \item \emph{Scalability}: The increased scalability of serverless platforms is advantageous for the application.
\end{itemize}

\begin{figure}[ht]
    \centering
    \includegraphics[width=0.8\linewidth]{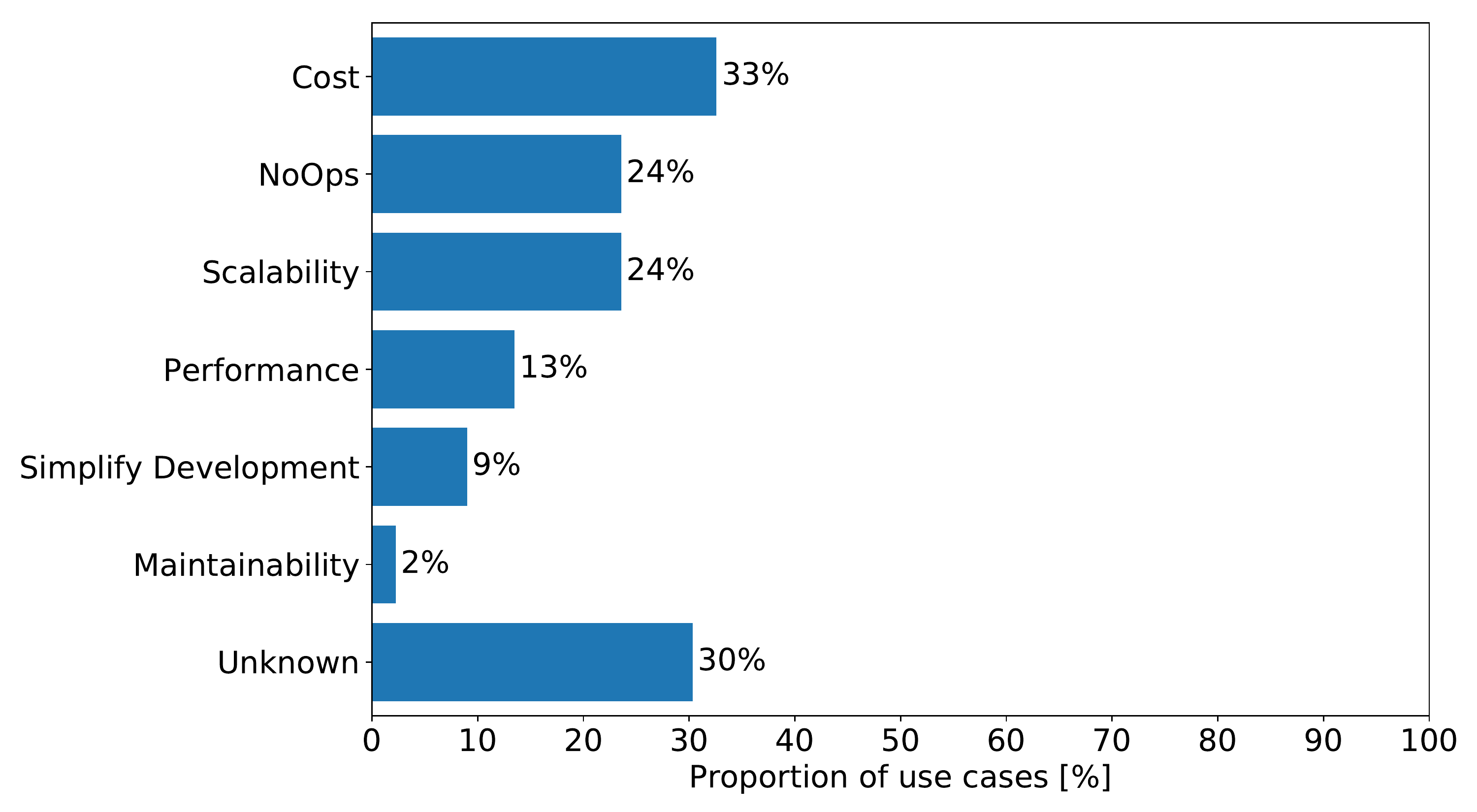}
    \caption{Distribution of the motivation behind adopting serverless among the surveyed use cases. Some use cases have multiple motivations.}
    \label{fig:motivation}
\end{figure}

\para{Results}
The results of our study can be observed in Figure~\ref{fig:motivation}. 
The biggest drivers for the adoption of serverless in our use cases are cost (33\%), the reduced operation effort (24\%), and the offered scalability (24\%).
Two further significant motivation behind the adoption seems to be the performance benefits (13\%) and the simplified development (9\%). 
However, the maintainability (2\%) only plays a minor role. 
For 30\% of the use cases, no specific motivation could be determined.\\

\para{Discussion} As the time savings by employing the NoOps paradigm of the serverless platforms can be converted to personnel costs, we observe that saving effort and costs seems to be a bigger contributor to the adoption of serverless than the offered performance and scalability improvements (although they are closely behind on second place).

It is important to note here, that there are many common pitfalls which can make serverless functions cost-inefficient. First, right now most providers bill by rounding up the execution time to the nearest 100ms. While this is negligible for most functions, this can be quite inefficient for very short-running functions. For example, if a functions runs for 10ms, it is billed for 100ms, which increases the billed duration tenfold. Secondly, most providers offer different function memory sizes and scale the other allocated resources such as CPU, I/O capacity and network bandwidth accordingly. A recent survey reports that about 50\% of serverless functions use the minimum size of 128MB~\cite{datadog}, which is reported to be inefficient for most serverless functions. Thirdly, at a very large scale, the raw infrastructure costs are significantly larger than for a traditional VM-based solution~\cite{eivy2017wary}. However, one might argue that the total cost of ownership could still be lower for the serverless solution due to the reduced operational overhead. In general, the the economic benefits of serverless computing heavily depend on the execution behavior and volumes of the application workload~\cite{eivy2017wary}.

\subsubsection{Cost/Performance Trade-off}
\label{sec:results:requirements:tradeoff}
\para{Description}
The cost/performance trade-off describes whether a use case tends to focus rather on cost optimization (i.e., \emph{cost-focused}) or rather on performance optimization (i.e., \emph{performance-focused}).
The trade-off is \emph{undefined} if cost and performance are equally important and \emph{unknown} if we could find no evidence towards any previously mentioned value in the provided use case description.\\

\begin{figure}[ht]
    \centering
    \includegraphics[width=0.8\linewidth]{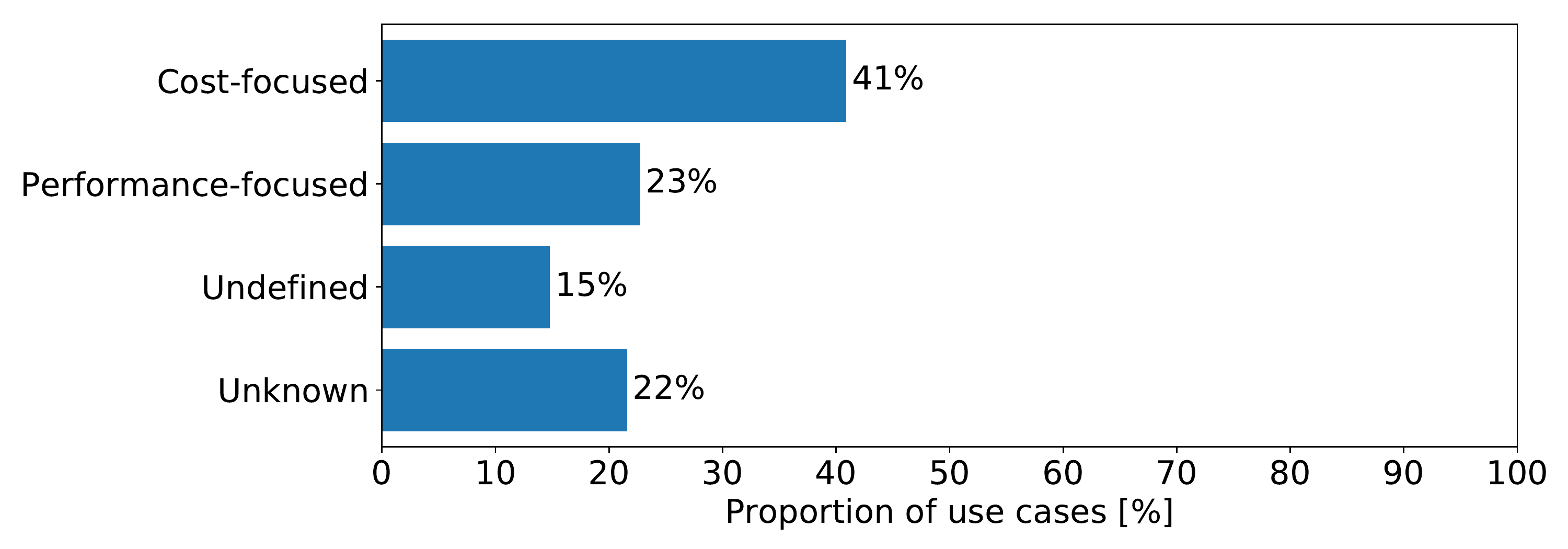}
    \caption{Distribution of the cost/performance trade-off among the surveyed use cases.}
    \label{fig:tradeoff}
\end{figure}

\para{Results}
Figure~\ref{fig:tradeoff} shows that cost is generally more important than performance for 41\% of the use cases.
Cost-focused use cases are also twice as common compared to performance-focused use case (23\%).
For 15\% of the use cases, cost and performance are equally important.
Finally, the trade-off remains unknown for 22\% of the use cases.\\

\para{Discussion}
The clear focus on cost optimization is plausible given that cost is a strong motivation for adopting serverless (see \Cref{sec:results:requirements:motivation}).
Serverless solutions, such as FaaS, were also associated with lower perceived total cost in another study~\cite{leitner2019MixedMethod}.

\subsubsection{Is Latency Relevant?}
\para{Description}
A diversity of use cases comes with a broad spectrum of expectations or even requirements on latency as a central performance metric. So we posed ourselves the question about the relevance of latency across the analysed serverless use cases. For this characteristic, we distinguish between four levels  plus \emph{Unknown}. The levels are:
\begin{itemize}
    \item \emph{Not important}: For these use cases we found evidence that latency does not play a central role. Delays and variations in latency are acceptable without disturbing the mode of operation.  
    \item \emph{For complete use case}: Latency plays a relevant role for the whole use case on a level of mostly unspecified expectations on latency and its variations over time, e.g., expected to exhibit a latency for convenient human user interaction.
    \item \emph{For parts of the use case}: The use case includes parts where latency is irrelevant and other parts where latency is of concern following the understanding of the level above as mostly unspecified expectation.  
    \item \emph{Real-time}: We select the level \emph{real-time} if evidence was found that there are soft latency requirements specified. Replies that take longer than a given upper time limit are becoming useless and are not further processed. This interpretation of \emph{real-time} is not implying safety critical states when latency requirements are violated. 
\end{itemize}

\begin{figure}[ht]
    \centering
    \includegraphics[width=0.8\linewidth]{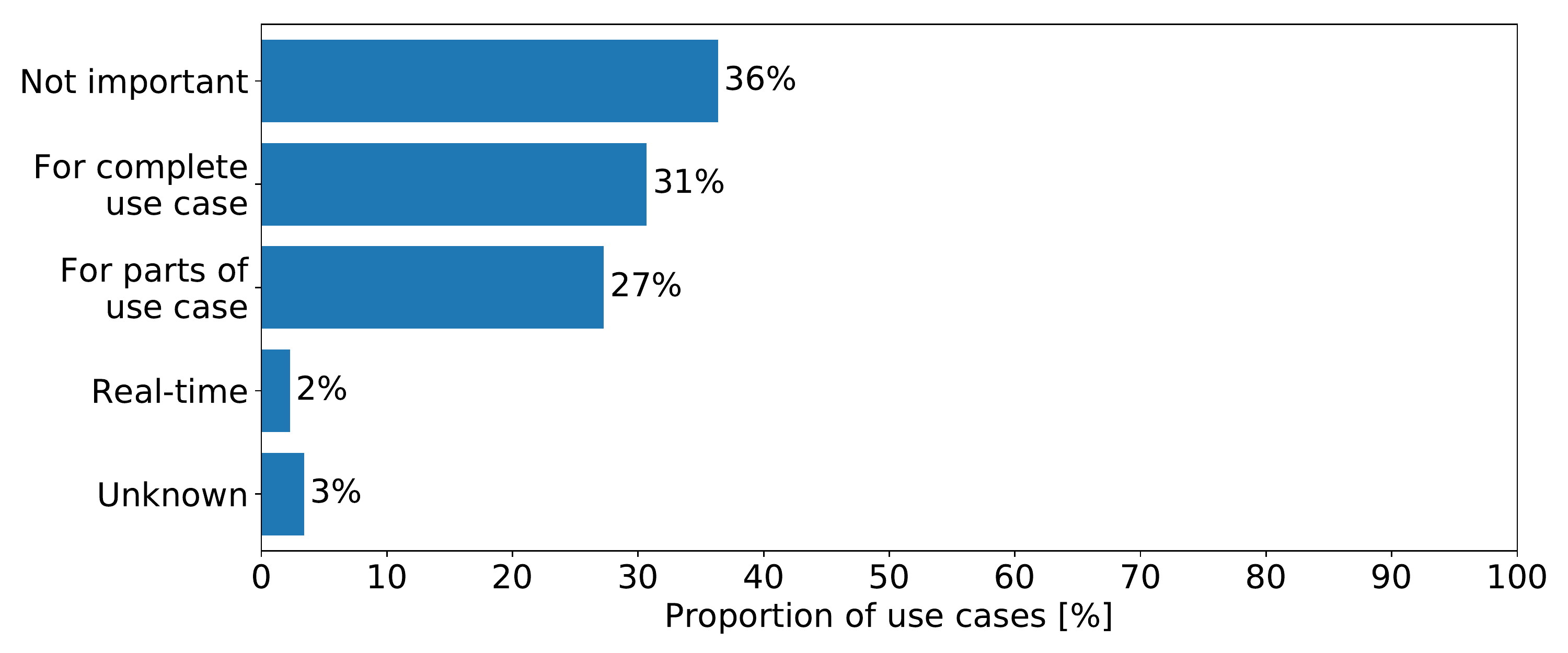}
    \caption{Distribution of latency importance among the surveyed use cases.}
    \label{fig:latency_relevant}
\end{figure}

\para{Results}
Figure~\ref{fig:latency_relevant} shows that in more than one third (36\%) of the use cases, latency does not play a role. On the other side, for 58\% of the analysed use cases, latency is of relevance (joining the respective levels). In 27\% this is only for parts of the use cases. 
The portion of use cases with real-time requirements on latency is small with only 2\%.
No clear assignment of was reasonable for 3\% of use cases. \\

\para{Discussion}
Serverless computing is especially convenient for triggered or scheduled background tasks that need to run from time to time without any latency requirements. But issues around function cold-starts and limited function life time have not shown to be a showstopper for serverless uses cases that expect a certain degree of stable latency, e.g. for a smooth interaction with human users. Also, over time, we would expect more examples for serverless use cases that come with stricter soft-real-time requirements as the platforms continue to mature. We doubt that serverless computing will accommodate use cases in production with hard real-time requirements and safety critical implications in case of violations.

\subsubsection{Locality Requirements}

\para{Description}
Migration of any application from dedicated servers in a possibly self-owned data center to a compute infrastructure managed by a cloud provider reduces the control over the locality where the code runs and data is persisted. From the early days of cloud computing on this remains still a possible issue or even show-stopper. 
We analyse the serverless use cases if requirements on locality are imposed. The reasons for locality requirements can differ broadly from regulatory to performance related ones.
Here, we distinguish between four levels of locality requirements plus the case ``unknown'': 

\begin{itemize}
    \item \emph{None}: There is evidence that for the given use case, no locality requirements are imposed.
    \item \emph{Multi-region}: The serverless application is or should be deployed in multiple regions, e.g. for improved latency or in tailored variations for specific geographic regions.
    \item \emph{Specific-region}: The serverless applications is required to run in a specific region.
    \item \emph{Edge}: The application or parts of it should run closer to a user or IoT device in an edge infrastructure
\end{itemize}

\begin{figure}[ht]
    \centering
    \includegraphics[width=0.8\linewidth]{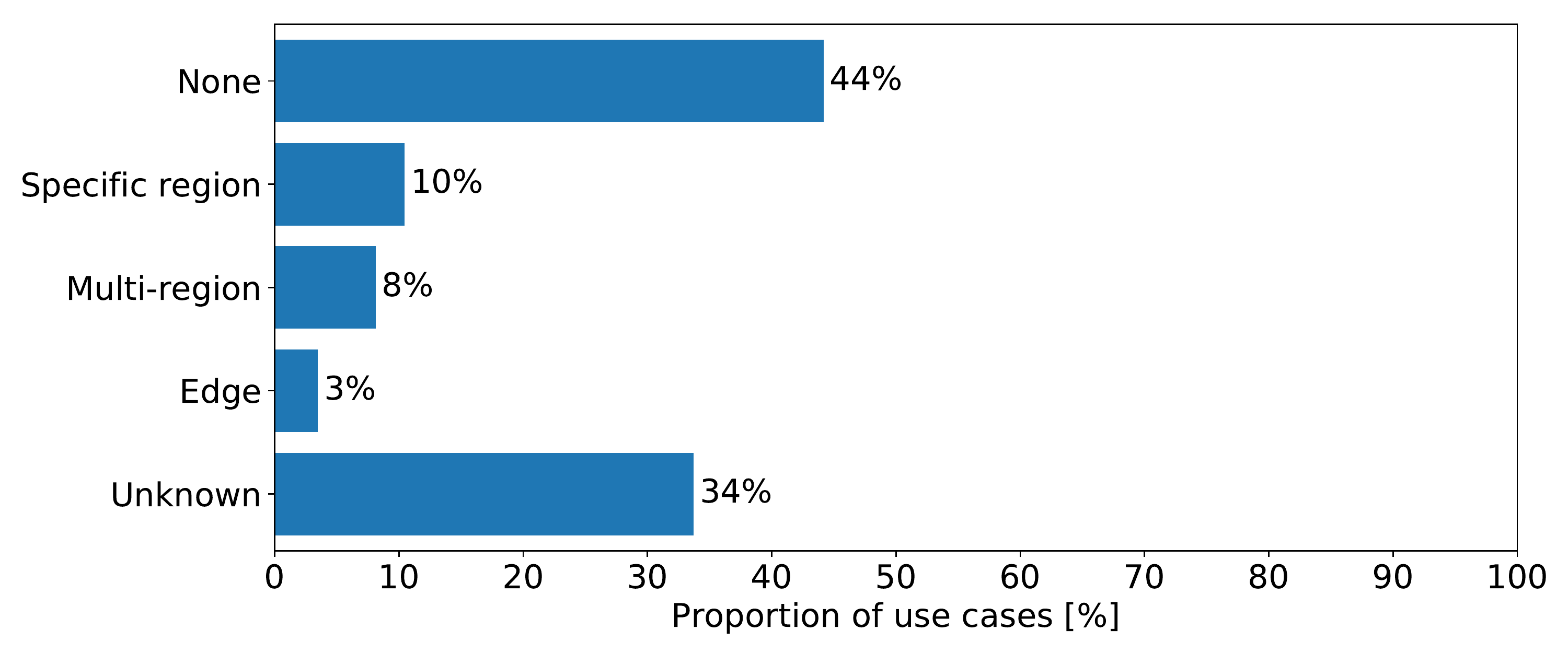}
    \caption{Locality requirement distribution among the surveyed use cases.}
    \label{fig:locality_requirement}
\end{figure}

\para{Results}
While we have unclear or unspecified locality requirements for 34\% of the use cases, Figure~\ref{fig:locality_requirement} shows that the biggest portion (44\%) comes with no locality requirements. For 21\% of the use cases, we found locality requirements. Out of those, 8\% are deployed across regions, while 10\% are run in specific regions. The remaining 3\% are tailored serverless solutions for edge computing. \\

\para{Discussion}
As serverless technologies and applications are maturing, we expect to see more business-critical elements of serverless applications in daily operation. The portion of serverless use-cases that comes with region-specific requirements will grow respectively. Furthermore, the use of serverless technologies for Edge computing can be seen as a trend of growing importance. Thus, we think it is likely to see a growing importance of the locality requirement ``Edge''. At the current time, for the dominating part of use cases, locality requirements are apparently not specified or not given yet. 

\subsection{Workflow Characteristics}
\label{sec:results:workflow}\label{sec:results:structural}

Many serverless use cases cannot use a single serverless function to meet their functional and non-functional requirements. 
Instead, such use cases require the execution of multiple functions, 
expressed and orchestrated as \textit{serverless workflows}. In this section, we investigate the characteristics of serverless workflows. However, not all use cases include workflows. Thus, we first investigate in  Section~\ref{sec:results:wf:is} which use cases are based on serverless workflows, and from then on we only report results for the use cases that do (the \textit{workflow use cases}).
\newline

\subsubsection{Is it a Workflow?}\label{sec:results:wf:is}
\begin{figure}[ht]
    \centering
    \includegraphics[width=0.8\linewidth]{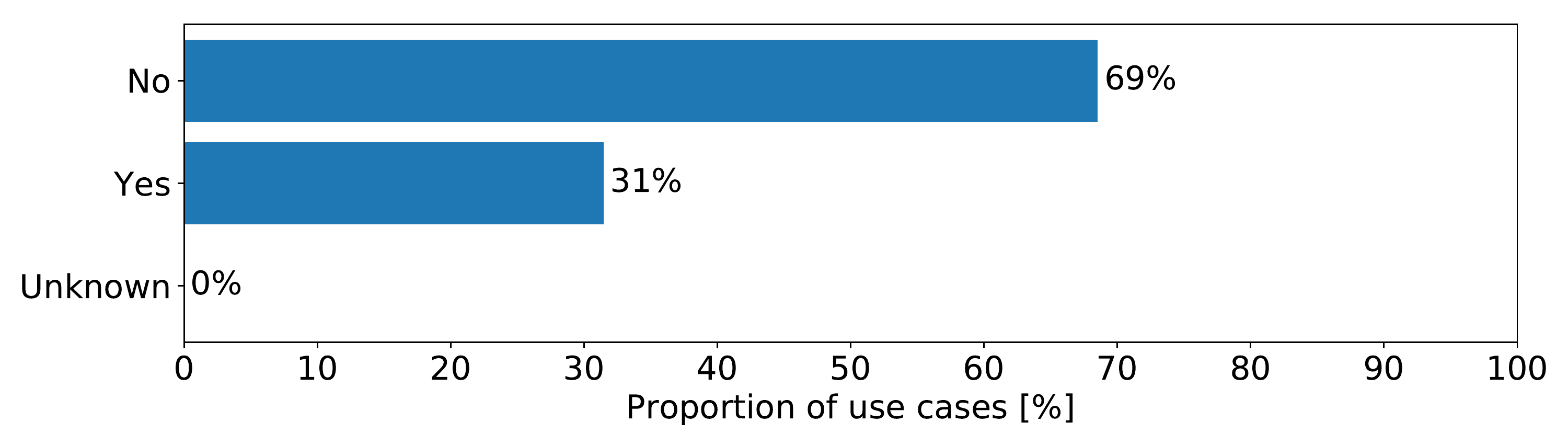}
    \caption{Percentage of use cases including workflows, among the surveyed use cases.}
    \label{fig:is_workflow}
\end{figure}

\para{Description}
We evaluate here the prevalence of serverless workflows among the surveyed use cases. A use case is categorized as a workflow (bar \textit{Yes} in Figure~\ref{fig:is_workflow}) if for a part or all of its functionality multiple serverless functions are needed. If not, the use case is not based on a workflow (\textit{No}). Use cases where this could not be determined are assigned \textit{Unknown}. \\

\para{Results}
As depicted in Figure~\ref{fig:is_workflow}, we observe that nearly a third (31\%) of the use cases include serverless workflows. The other use cases (69\%) are simple enough that one or a couple of independent serverless functions can fully provide the desired functionality. No use case was labeled Unknown.\\

\para{Discussion}
The relative prevalence of workflows in use cases is important, as it hints that serverless use cases are getting more and more complex. The evolution of use cases in fields such as grid computing and more recently cloud computing is indicative that, once workflows become acceptable practice, they become increasingly more prevalent~\cite{hey2009fourth,isom2012your,deelman2018future}. Interesting too is the lack of use cases categorized as Unknown, which indicates that the presence or absence of workflows for any relevant use case is one of the clearest questions to answer. 

\subsubsection{Workflow Coordination}\label{sec:results:wf:coord}

\para{Description}
The use of workflows currently does not constrain the method used for orchestration. There are various approaches to ensure that tasks---or serverless functions---are executed in a coordinated way, e.g., functions can use events to trigger the start of a new task or for other purposes, a task can act as a coordinator for a specific workflow structure, or a workflow engine can orchestrate arbitrary workflows. To evaluate which of these approaches is prevalent, we surveyed their use across workflow use cases. More formally, for this part we categorized orchestration techniques as follows: 

\begin{enumerate}
    \item \emph{Event} groups all use cases that rely on event-driven orchestration. In this approach, workflows are constructed by configuring functions to be triggered to execute on the arrival of the completion (or failure) events of other functions. Typically, this method requires functions to explicitly listen for and publish their results or errors to a message queue---though in some cases platforms this functionality is built-in and no explicit interaction with an external message queue is needed.  
    
    \item \emph{Local coordinator} groups all the applications which rely on programmed, user-side logic to take care of the orchestration. An application running on the user's machine, such as a GUI or the client-side JavaScript running on a web page, invokes the functions in the appropriate order, and ensures that each function is executed with the correct configuration and input data.
    
    \item \emph{Workflow engine} contains the use cases that delegate the coordination to a dedicated workflow management system. This workflow engine has functionality to ensure the correct orchestration, along with higher-order concerns, such as (data) provenance, monitorability, and task scheduling optimizations. The workflows typically need to be specified in a consistent format using a set of workflow primitives that are supported by the workflow engine. Compared to the local coordinator, a workflow engine can also be seen as an external, persistent coordinator.
    
    \item \emph{Unknown} captures the use cases where we could not determine the coordination approach, for example because of lacking or lack of documentation.
\end{enumerate}

\begin{figure}[ht]
    \centering
    \includegraphics[width=0.8\linewidth]{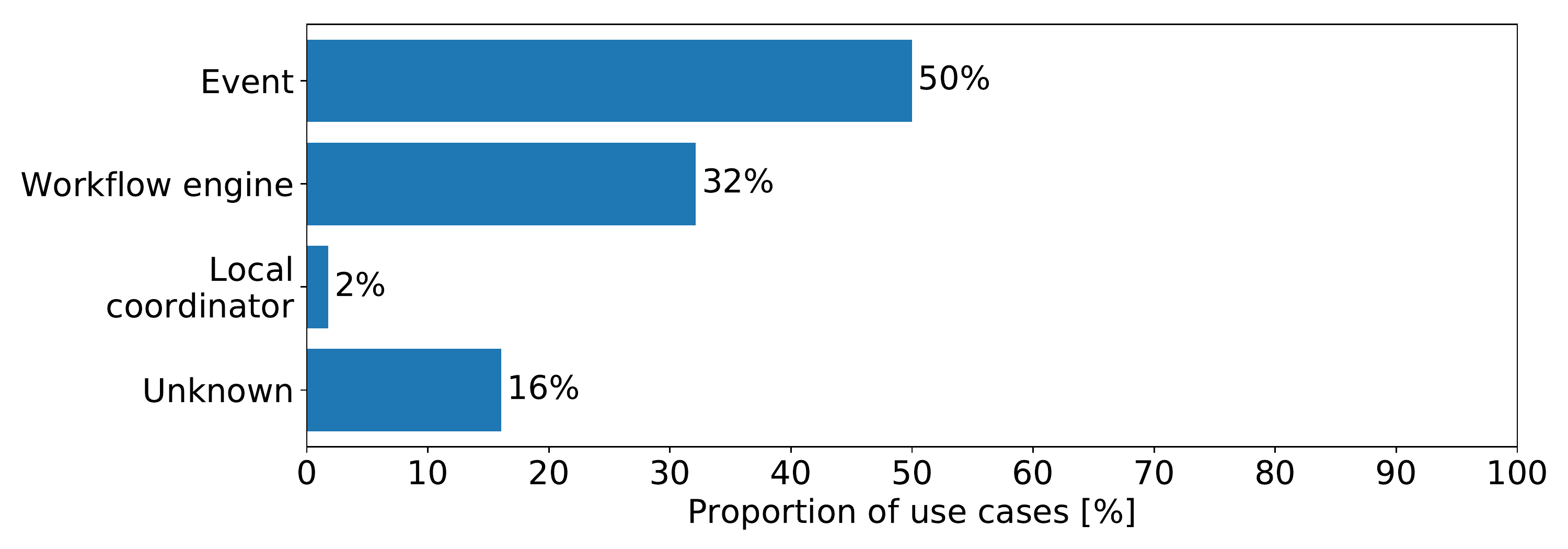}
    \caption{Distribution of workflow coordination approach among the surveyed use cases.}
    \label{fig:coordination}
\end{figure}

\para{Results}
As depicted in Figure~\ref{fig:coordination}, half of the serverless workflows rely on events for the coordination (50\%). Slightly less prevalent, about one-third (32\%) of the use cases rely on a dedicated workflow engine to ensure correct coordination. Only a few (2\%) defer the coordination to a local coordinator. For 16\% of the workflow-based use cases the approach could not be determined.\\

\para{Discussion}
Our results indicate that event-driven workflows are currently most prevalent. Inspecting the individual use cases, we find that this is in part caused by implicit workflows; use cases that do not explicitly construct workflows, but instead configure the function triggers in such a way that these form simple pipelines. While this approach can address simple workflows and especially chains of a few tasks,
experience from the fields of grid and cloud computing indicates using this approach will not scale to future workflows. We further find that, as the workflows grow in complexity, workflow engines are more often used to coordinate the workflows. In contrast to cloud-side coordination techniques, the use of local coordinators is unpopular because it distributes to the user more complex logic and, in part, because it is difficult to maintain. Furthermore, such an approach could be less reliable in operation -- cloud-side coordination techniques and workflow engines are carefully engineered for fault-tolerance, which significantly exceeds the typical development effort of local coordinators.

\subsubsection{Workflow Structure}\label{sec:results:wf:structure}

\para{Description}
The complexity of a workflow is mostly determined by its structure. A {\it bag of tasks} is a simple workflow (in mathematical terms: a degenerate workflow), which consists of a set of tasks that can be executed in any arbitrary order. 
Another common workflow structure is the {\it sequential workflow}, where all tasks need to be executed sequentially. 
We further define {\it complex workflows} as workflows that include significantly more complex structure than the previous types, including (multi-stage) gather and scatter operations, workflows with conditional execution of (some) tasks, workflows with loops, and fully dynamic workflows.\\

\begin{figure}[ht]
    \centering
    \includegraphics[width=0.8\linewidth]{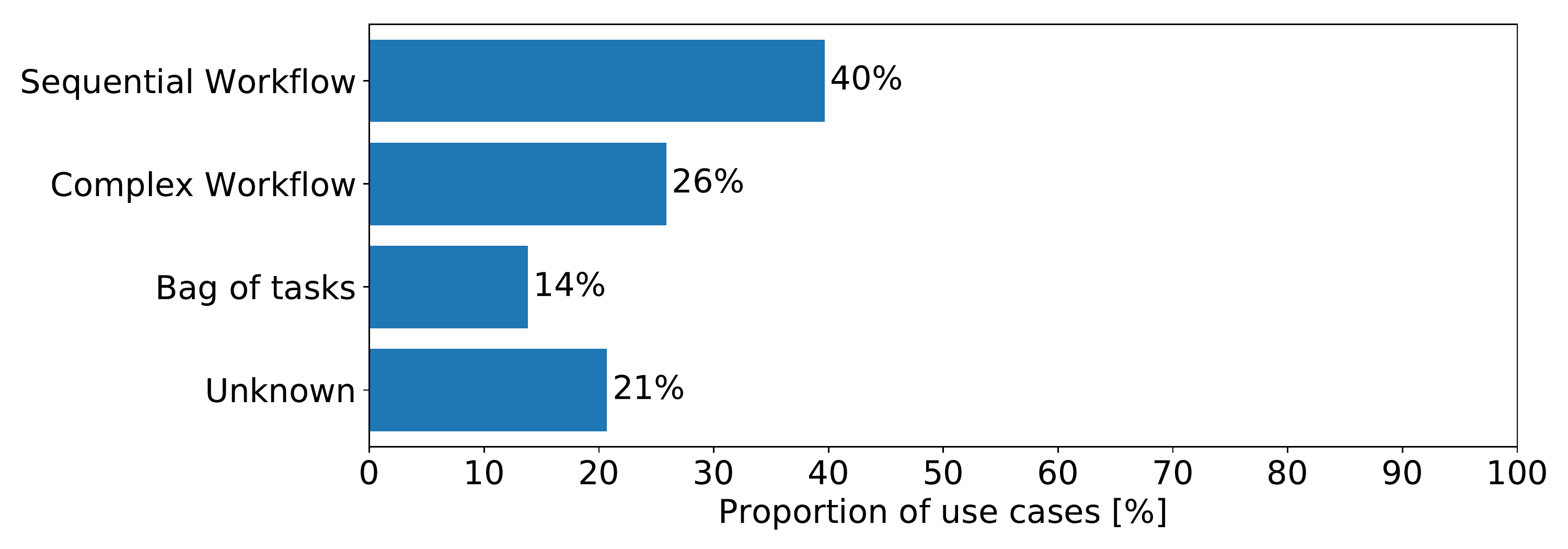}
    \caption{Distribution of the workflow structure among the surveyed use cases.}
    \label{fig:workflow_structure}
\end{figure}

\para{Results}
Figure~\ref{fig:workflow_structure} depicts the results.
Sequential workflows are the most popular workflow structure for serverless applications, with 40\% of the workflow use cases including them. 
Including sequential workflows and bags of tasks (14\%), over half (54\%) of the workflow use cases only include non-complex workflows. 
About one-quarter (26\%) of the serverless applications are complex workflows. 
Last, for over one-fifth (21\%) of the workflow use cases we were not able to determine the workflow structure.\\

\para{Discussion}
For serverless applications, simple workflow structures (bag of tasks and sequential workflows) are more than twice as common as more complex workflow structures. We hypothesize that this is because serverless applications are currently mostly used for comparatively simple tasks and rarely for complex data analysis.
Another possible explanation is the lack of workflow engines (Section~\ref{sec:results:wf:coord}); it is difficult to orchestrate arbitrarily complex workflows without such an engine. \\

The lack of bags of tasks can probably attributed to the fact that serverless applications come with built-in scalability when the functions can be conveniently executed in parallel. For example, resizing a collection of images can be conveniently implemented as a bag of many tasks, where each task invokes the same image-resizing function. In this case, the serverless platform has the capability to execute this case, without further need for orchestration from the user.

\subsubsection{Workflow Size} \label{sec:results:wf:size}

\para{Description} Next, we study workflow size, expressed as the number of tasks in the workflow. 
We aggregate all use cases into three groups:
\begin{enumerate}
    \item \textit{Small workflows}, containing 2--10 functions,
    \item \textit{Medium-size workflows}, invoking 10--1000 functions, and
    \item \textit{Large workflows}, comprised of more than 1000 function invocations. 
\end{enumerate}

\begin{figure}[ht]
    \centering
    \includegraphics[width=0.8\linewidth]{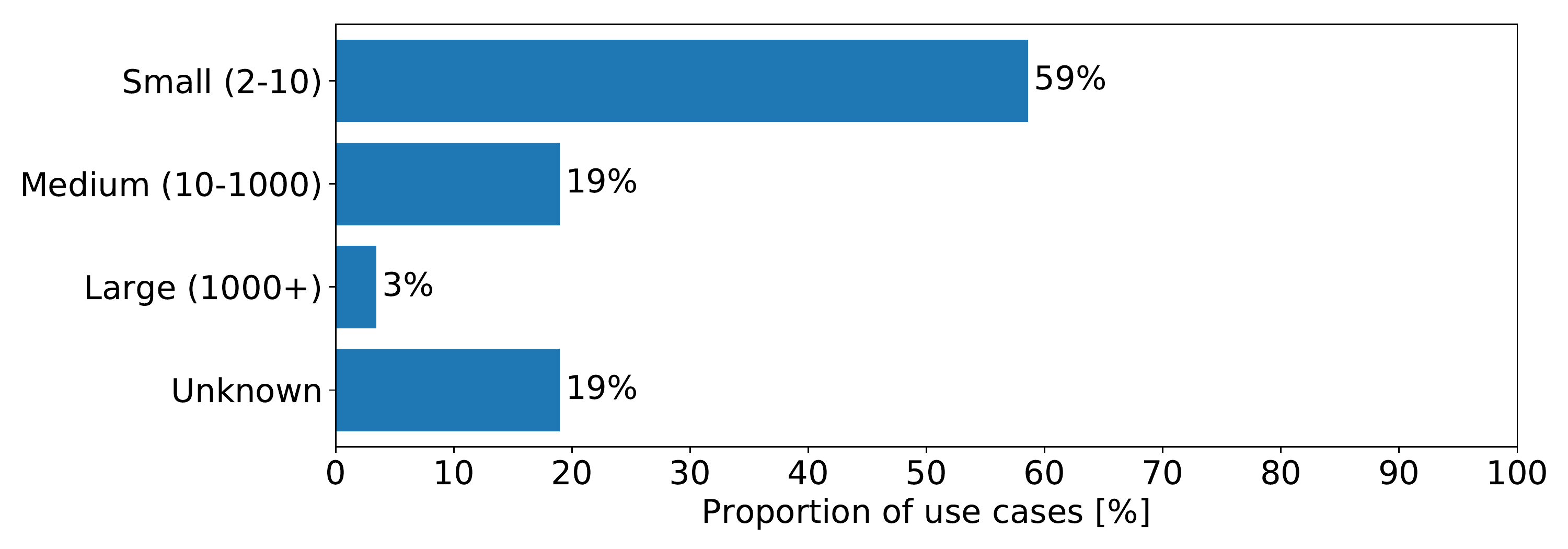}
    \caption{Workflow size distribution among the surveyed use cases.}
    \label{fig:workflow_size}
\end{figure}

\para{Results} Figure~\ref{fig:workflow_size} depicts the results of our analysis. 
The majority (59\%) of analyzed workflows are small workflows.
Around one fifth of use cases (19\%) are medium-size, and only few (3\%) qualify as large workflows.
Nearly one-fifth of the workflows (19\%) could not be assigned to one of the groups. \\

\para{Discussion} Our results suggest that a majority of workflow executions are small; because these workflows are only composed of ten or less individual function executions, they are also relatively short-lived. This is consistent with the characteristics of early workflows in engineering and in scientific prototypes~\cite{conf/cgiw/OstermannIPFE08}. Only about one-fifth of the workflows are medium or large-sized. Similarly to the previous section, we hypothesize that orchestrating workflows of this size is dependent on the presence of an automated facility, such as a workflow engine. (The other hypothesis introduced in Section, that serverless workflows are currently used for relatively simpler tasks, does not limit the size of the workflow -- in the earlier example, the image-resizing workflow can run 10,000s or even 100,000s of functions~\cite{bharathi2008characterization, deelman2018future}.)

\subsubsection{Workflow Internal Parallelism}

\para{Description}
For those use cases in which a workflow of serverless functions was present, we further analyzed whether they present \emph{internal parallelism}---at least an instance of multiple functions running in parallel---or not. \\

\begin{figure}[ht]
    \centering
    \includegraphics[width=0.8\linewidth]{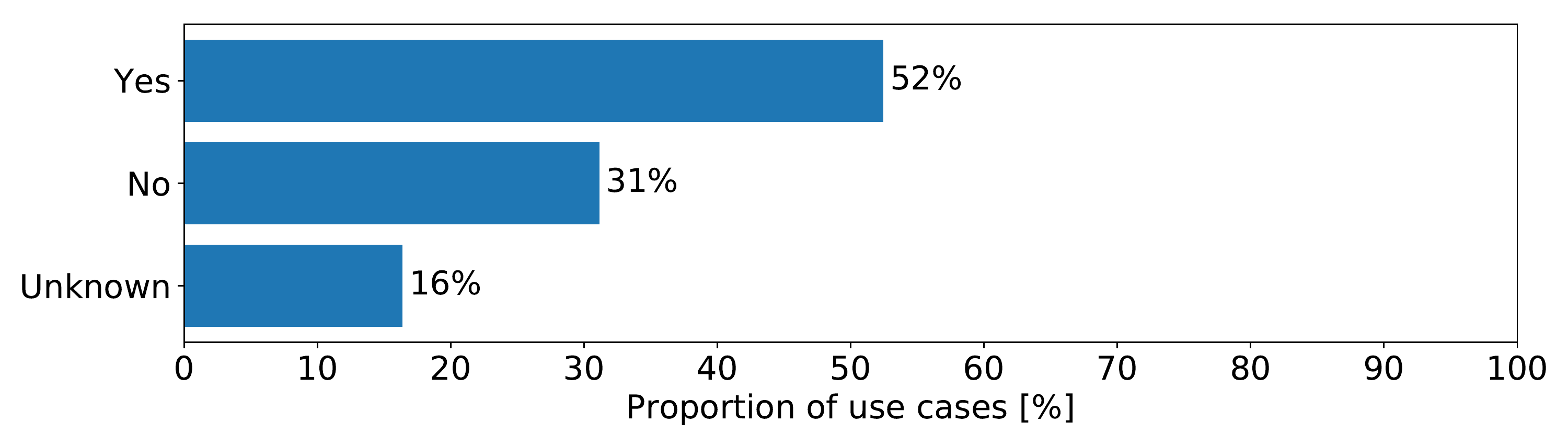}
    \caption{Percentage of workflows with internal parallelism among the surveyed use cases.}
    \label{fig:parallelism}
\end{figure}

\para{Results}
From Figure~\ref{fig:parallelism}, we observe that most workflows (52\%) exhibit internal parallelism; about one-third (31\%) of the workflows are simpler, exhibiting no internal parallelism.
We could not obtain this information (\emph{unknown}) for 16\% of the workflows. \\

\para{Discussion}
The high prevalence of workflows with at least some level of internal parallelism calls for workflow managers that are native to---or well integrated with---the serverless framework, to facilitate workflow composition and management yet deliver parallelism with low overhead.
\\

\section{Threats to Validity}
\label{sec:validity}\label{sec:threats}
We discuss potential threats to validity and mitigation strategies for internal validity, construct validity, and external validity.

\subsection{Internal Validity}
Manual data extraction can lead to inaccurate or incomplete data.
To mitigate this threat, we established and discussed a review protocol prior to reviewing, continuously discussed upcoming questions during the review process, and performed redundant reviews through multiple reviewers.
In our review protocol, we established an exhaustive list of potential values for each characteristic and configured automated validation, which immediately highlighted deviations from these values.
For characteristics with thematic coding, we continuously refined their values in regular meetings during the review process.
To address potential individual bias, we performed two independent reviews for each use case, quantified the inter-rater agreement after an initial review round through Fleiss' Kappa, and resolved each disagreement in an extended discussion and consolidation phase.

\subsection{Construct Validity}
To align the goal of this study (i.e., comprehensive understanding of existing serverless use cases) with the data extraction, we compiled a list of 24 characteristics covering 5 different aspect groups.
We conducted and discussed this selection process together in an international working group with authors from 5 different institutions but other researchers might consider different characteristics as relevant.

\subsection{External Validity}
Our study was designed to cover use cases from open source projects, white literature, and grey literature but we cannot claim generalizability to all serverless use cases.
For open source projects, we filtered non-trivial projects from the most popular open source repository (i.e., GitHub) but might have missed projects published in other repositories.
However, we are unaware of such other repositories and also did not discover any among our other use cases from white and grey literature.
Our white literature collection is based on a curated dataset on serverless literature and complemented with articles known to the authors but we might have missed more recent articles uncovered in the dataset and unknown to all authors.
Grey literature use cases mostly focus on provider-reported case studies, an existing collection of grey literature use cases, and sources known to the authors.
We only partially cover corporate use cases as many of them remain unpublished and others provide insufficient details to conduct a meaningful review, which is similar to FaaS platforms~\cite{DBLP:journals/internet/EykIGEBVTSHA19}.
Our scientific computing use cases are limited to the aerospace domain originating from a national aerospace institution.

\section{Conclusion and Future Work}
\label{sec:conclusion}\label{sec:futurework}

The emergence of serverless computing has already led to a diverse design space, with tens of serverless platforms and the participation of all major cloud providers.
We identify in this work the need for a systematic, comprehensive study of serverless use cases, which could help the development of serverless techniques and solutions in the fields of software engineering, distributed systems, and performance engineering. \\

We have proposed a systematic process to identify, collect, and characterize serverless use cases.
To identify use cases, the process considers open-source software projects, peer-reviewed literature, self-published material, and domain knowledge.
To collect the use cases, the process proposes a structured repository, from which reviewers take and characterize each use case alongside 24 features of interest.
Each use case is covered by the following types of features:
(a) \textit{general} characteristics, such as platform, application type and domain, whether the use case was observed in production, and whether the use case provides open-source software;
(b) \textit{workload} characteristics, such as the execution pattern, burstiness, types of triggers, and data volume;
(c) \textit{application} characteristics, such as programming language(s) used to develop it, the resource bounds, whether the application depends on external services, etc.;
(d) the \textit{requirements} posed by the use case, such as locality and latency, or the performance-cost trade-off; and
(e) \textit{workflow} characteristics, including structure, size, and internal parallelism. \\

Using this process, we have collected and characterized a total of \totalusecases serverless use cases from four different sources. Our systematic and comprehensive study reveals that:
\begin{enumerate}

    \item We find a dominating portion of serverless use cases already being in production with AWS as the most popular platform and web services being the most common application domain. 

    \item Serverless workloads tend to exhibit on-demand execution patterns exemplified by 81\% bursty workloads, which makes their load hard to predict.
    
    \item Most cloud functions (67\%) are short-running, with running times in the order of milliseconds or seconds, thus requiring serverless frameworks that impose small overheads when running functions.
    
    \item Cost savings (both in terms of infrastructure and operation costs) are a bigger driver for the adoption of cost than the offered performance and scalability gains.
    
    \item 
    We observe an increasing trend toward ever-larger, ever more complex workflows, indicating the need to move toward (cloud-native) workflow engines.
\end{enumerate}

Last, but not least, we see this study as a step toward a community-wide policy of sharing and discussing about use cases. Persisting beyond our effort, such use cases could stimulate a new wave of serverless designs, facilitate meaningful tuning and benchmarking, and overall prove useful for both academia and industry.
We therefore extend an open invitation to prospective new collaborators in the SPEC-RG Cloud group.

\cleardoublepage
\pagenumbering{gobble}
\pagestyle{empty}
\renewcommand\bibname{References}

\bibliographystyle{IEEEtranSA}
\bibliography{maindoc}

\clearpage
\appendix
\appendixpage
\addappheadtotoc
\section{Scientific Use Cases}
\label{appendix}

\subsection{Copernicus Sentinel-1 for near-real time water monitoring}

\vspace{1px}
\hspace{1cm} \textit{Contact persons:} Nico Mandery, Torsten Riedlinger, Maximilian Schwinger \\
\vspace{-5px}

Floods occur frequently and across most regions around the globe. They affect lives, infrastructures, economics and local ecosystems. The economic consequences of flood damage impacts the most vulnerable members of society disproportionately. Emergency responders often request Earth Observation based crisis information for flood monitoring to target the frequently limited resources and to prioritize response actions during a disaster situation.\\

The European Earth Observation program COPERNICUS is providing satellite data and products suitable for a variety of environmental and security applications. The Sentinel-1 radar satellites can be used for various applications in the field of marine and land surface dynamics, e.g. for the detection of water bodies, the mapping of its seasonal dynamics and for the monitoring of flood events. On average, Sentinel-1 provides approximately 1200 scenes per day, creating a daily amount of data in the order of 800 GB.\\

This application facilitates the near-real time detection and monitoring of flooded areas and can therefore provide vital information for decision makers, scientists and the general public.\\

\begin{figure}[ht]
    \centering
    \includegraphics[width=0.3\linewidth]{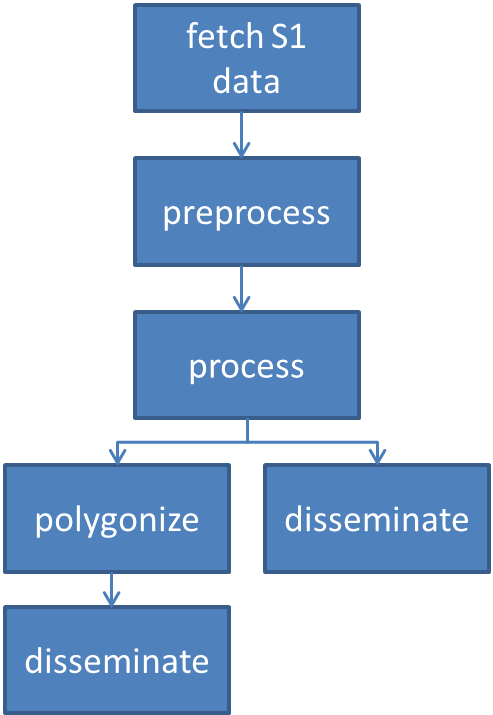}
    \caption{Step dependencies during extraction of information from Sentinel 1 data.}
    \label{fig:scientific1}
\end{figure}
A near-real time monitoring system for the provision of Sentinel-1 based products was developed at DLR-DFD in the last years (Martinis et al. 2018, Twele et al. 2016), which is continuously improved in terms of thematic quality and processing capacity.\\

The radar-based processing chains make use of an automatic hierarchical tile-based thresholding approach in combination with fuzzy-logic-based post-processing for the unsupervised extraction of the flood extent (Martinis et al. 2018). The processing chain can be divided into pre-processing (internal calibration and terrain correction) and the thematic processing (derivation of water bodies, dynamic and flood extent). The results are disseminated as raster and vector files, including the provision through web-services.\\

\begin{figure}[ht]
    \centering
    \includegraphics[width=0.9\linewidth]{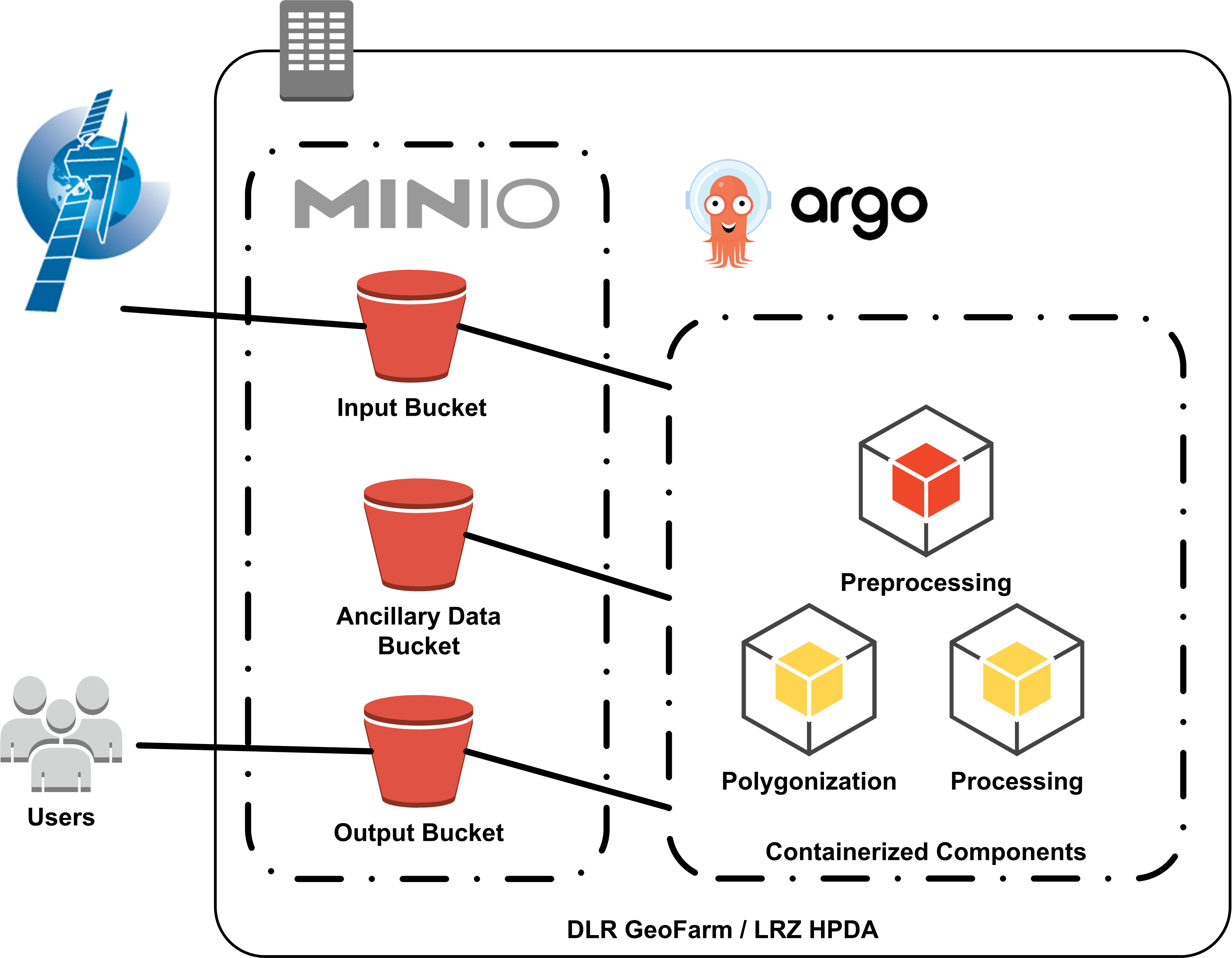}
    \caption{Architecture Overview of the S1 water monitoring system implemented at DLR.}
    \label{fig:scientific2}
\end{figure}

The processing is performed in a cloud native environment. Argo Workflows are orchestrating docker container native workflows in a Kubernetes environment. Using Kubernetes and Argo allows an easy evolution of the processing system and the implementation and   adapting of other thematic processing tasks in the future.\\
The physical infrastructure currently used for the Sentinel-1 application is the DLR-DFD multi-purpose processing Environment GeoFarm. The near-real time application is planned to be ported to DLRs new High performance data analytics (HPDA) -Infrastructure at the Leibnitz Rechenzentrum (LRZ) in Munich.\\

\textit{References:\\
Martinis, S., S. Plank, and K. Cwik, “The use of Sentinel-1 time-Series data to improve flood Monitoring in arid areas”, Remote Sensing, vol. 10 (582), pp. 1-13, 2018\\
Twele, A., W. Cao, S. Plank, and S. Martinis, “Sentinel-1 based flood mapping: a fully automated processing chain,” International Journal of Remote Sensing, 2016, vol. 37, no. 13, pp. 2990-3004, 2016}

\clearpage
\subsection{Terra\_Byte - High Performance Data Analytic for Earth Observation}

\vspace{1px}
\hspace{1cm} \textit{Contact person:} Maximilian Schwinger \\
\vspace{-5px}

Under the cooperation agreement Terra\_Byte between Europe’s second largest high performance computing center LRZ (Leibnitz Super Computing Center) and the DLR (German Aerospace Center) a high performance data analytic infrastructure fit for the specific requirements of earth observation processing is procured and developed. The environment is composed from an optimized hardware layer which can answer the high I/O requirements of earth observation requirements and a software stack, which provides earth observation scientists an easy access to the available resources. \\

Core of the HPDA Terra\_Byte is a large high performance online storage based on the LRZ’s Data Science Storage (DSS) concept which provides more than 30 PByte of relevant earth observation data online for use in different applications. The identification and access of the data is one of the major tasks to be accomplished: applications need a simple way to identify relevant data in hundreds of millions of data files, each larger than 1GByte.
Besides  identification and access of data,  usage of the significant computing infrastructure of HPDA-Terra\_Byte with simple mechanisms matching the needs of earth observation scientists are required. To provide this easy access and simple scalability the HPDA infrastructures stack will provide Platform as a service (PaaS) as well as Function as a service (FaaS) capabilities.\\

\begin{figure}[ht]
    \centering
    \includegraphics[width=0.8\linewidth]{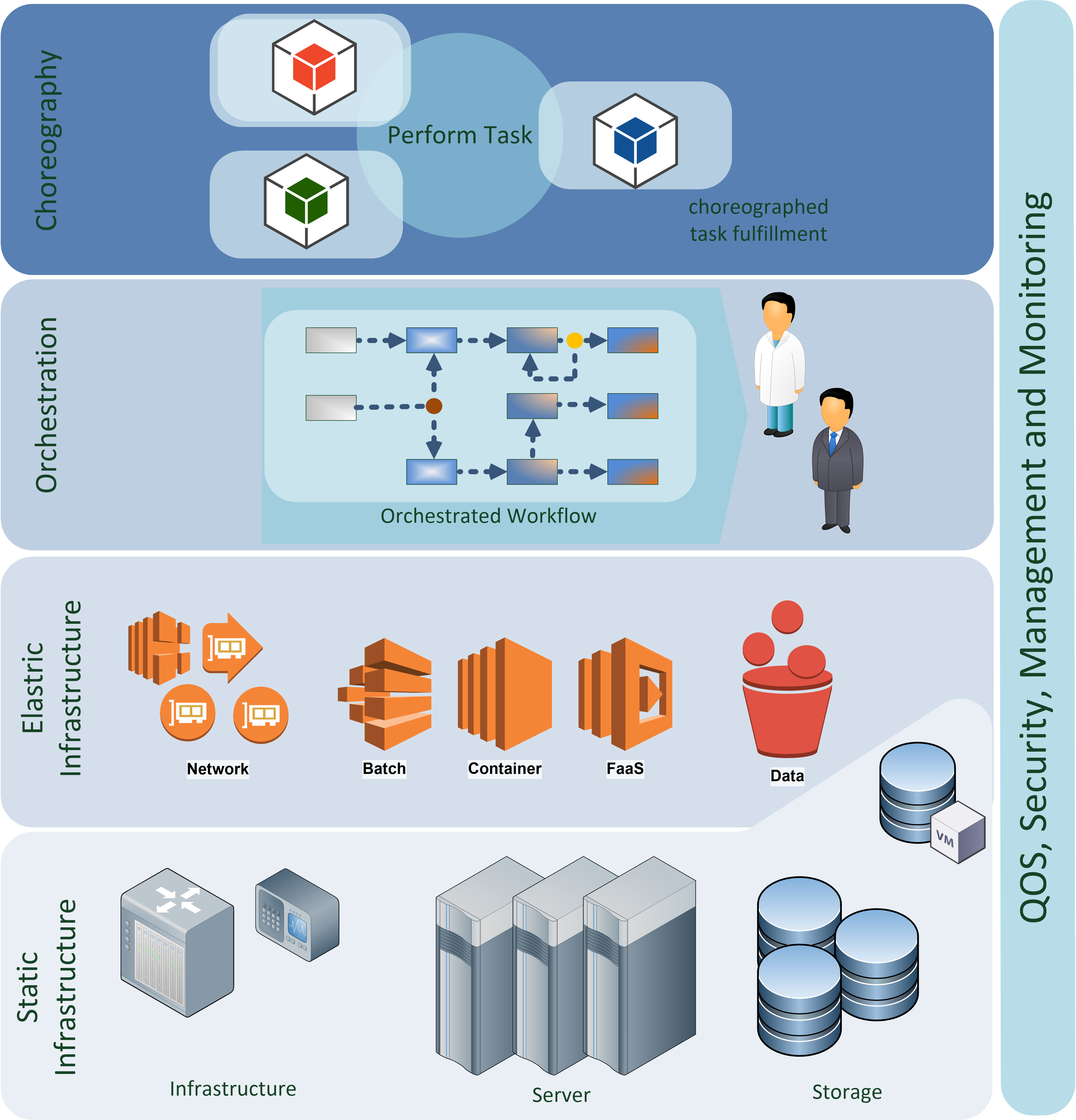}
    \caption{Overview of the intended implementation of HPDA Terra\_Byte.}
    \label{fig:scientific6}
\end{figure}

To achieve this target an analysis and extension of available open source software will take place. For the usage within the infrastructure a set of functions will be developed taking away logistical core tasks from the scientists to provide them the freedom to focus on their core work.
\subsection{Reprocessing Sentinel 5 Precursor Data with ProsEO}

\vspace{1px}
\hspace{1cm} \textit{Contact person:} Maximilian Schwinger \\
\vspace{-5px}

The European Copernicus program provides a variety of satellite data products fit for applications in monitoring of environment and security. The Sentinel 5 precursor, carrying the hyperspectral instrument TROPOMI, continuously monitors the earth's atmospheric composition. Data is produced continuously but progress in understanding of the data as well as the earth's atmosphere leads to a continuous development in the algorithms retrieving trace gas concentration from the sensor measurements. Due to this progress a reprocessing of the whole missions data is required from time to time. The amount of input here is in a petabyte scale, the number of files to be processed is in the scale of a million files and the time available is weeks. \\

The reprocessing system proposed has to follow a complex dependency graph of trace gas and cope with a multitude of configuration dependencies between different processor versions and configurations. The combination of processor version and configuration defines the requirement of an input product with a specific processor version and configuration produced. The processing of a product is triggered by a simple request of a specific product produced by a version/configuration of a processor. The dependency graph is generated by prosEO and the triggering of functions on hardware is done by Kubernetes. The processors are encapsulated in docker images. \\

\begin{figure}[ht]
    \centering
    \includegraphics[width=\linewidth]{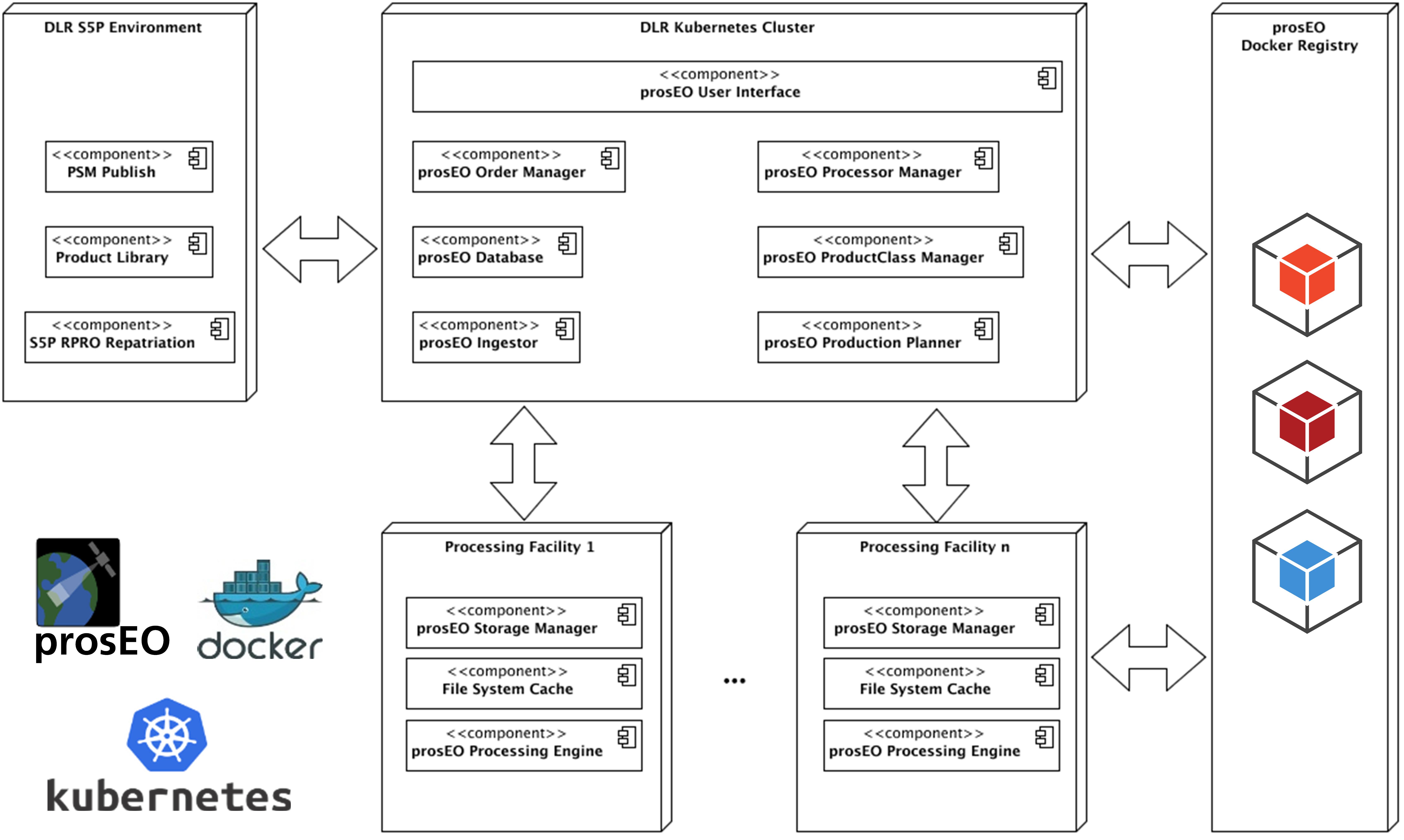}
    \caption{Architectural Overview of ProsEO - a framework for generating higher level products from satellite data.}
    \label{fig:scientific4}
\end{figure}

prosEO is capable of handling multiple cloud providers (processing facilities) depending on cost and data availability decisions.

\subsection{Tandem-L exploitation platform}

\vspace{1px}
\hspace{1cm} \textit{Contact persons:} Wolfgang Balzer, Maximilian Schwinger \\
\vspace{-5px}

The Tandem-L mission is a German L-band Radar project proposal currently in ECSS phase B of its development cycle. Tandem-L is a constellation of two satellites capable of observing dynamic processes on the earth’s surface in unpreceded quality.\\

Tandem-L will generate a daily amount of 8 TByte of raw acquisition data which is processed on ground to 13 interdepending products, operationally. In addition to this it is possible for scientists to implement an own processor which then can be interconnected with the operational processing streams. As no operational knowledge on processing streams, hardware allocation and interdependencies will be handed to external scientists developing processors the processor will be added to the exploitation platform as an additional function with interdependencies to other functions of the exploitation platform.

\subsection{Global Urban Footprint}

\vspace{1px}
\hspace{1cm} \textit{Contact persons:} Thomas Esch, Maximilian Schwinger

\hspace{1cm} \textit{Source:} \url{https://www.dlr.de/guf}\\
\vspace{-5px}

Currently, more than half of the world’s population are urban dwellers and this number is still rapidly increasing. Since settlements---and urban areas in particular---represent the centers of human activity, the environmental, economic, political, societal and cultural impacts of urbanization are far-reaching. They include negative aspects like the loss of natural habitats, biodiversity and fertile soils, climate impacts, waste, pollution, crime, social conflicts or transportation and traffic problems, making urbanization to one of the most pressing global challenges. Accordingly, a profound understanding of the global spatial distribution and evolution of human settlements constitutes a key element in envisaging strategies to assure sustainable development of urban and rural settlements. \\

In this framework, the objective of the “Global Urban Footprint” (GUF) project is the worldwide mapping of settlements with unprecedented spatial resolution of 0.4 arcsec (\textasciitilde 12 m). A total of 180 000 TerraSAR-X and TanDEM-X scenes have been processed to create the GUF. The resulting map shows the Earth in three colors only: black for “urban areas”, white for “land surface” and grey for “water”. This reduction emphasizes the settlement patterns and allows for the analysis of urban structures, and hence the proportion of settled areas, the regional population distribution and the arrangement of rural and urban areas.
 When looking at the entire GUF at once, mainly the metropolitan regions in Europe, USA East Coast and Asia stand out. When focusing on the full resolution of the GUF, one can even recognize small villages stretched along roads, single farm houses or non-built-up corridors in megacities. For a comprehensive and objective analysis of the settlement patterns, the DLR additionally developed an approach to display the spatial networks between the mapped settlements. This enables the computation of various form and centrality measures, which are used for qualitative and quantitative characterization of settlement patterns at different spatial units ranging from global to local scale.\\

The GUF exhibits a high potential to enhance climate modelling, risk analyses in earthquake or tsunami regions and the monitoring of human impact on ecosystems. Moreover, the it also can be employer as basis for monitoring both the historical growth of different settlements, as well as their ongoing and future development.. This will allow effective comparative analyses of urban dynamics among different regions of the world.

\subsection{DESY - High Throughput Data Taking}

\vspace{1px}
\hspace{1cm} \textit{Contact persons:} Patrick Fuhrmann, Michael Schuh, Maximilian Schwinger

\hspace{1cm} \textit{Source:} \url{https://www.desy.de/about_desy/desy/index_eng.html}\\
\vspace{-5px}

DESY is one of the world’s leading accelerator centres. Researchers use the large-scale facilities at DESY to explore the microcosm in all its variety---from the interactions of tiny elementary particles and the behaviour of new types of nanomaterials to biomolecular processes that are essential to life. The accelerators and detectors that DESY develops and builds are unique research tools. The facilities generate the world’s most intense X-ray light, accelerate particles to record energies and open completely new windows onto the universe. That makes DESY not only a magnet for more than 3000 guest researchers from over 40 countries every year, but also a coveted partner for national and international cooperations. Committed young researchers find an exciting interdisciplinary setting at DESY. The research centre offers specialized training for a large number of professions. DESY cooperates with industry and business to promote new technologies that will benefit society and encourage innovations. This also benefits the metropolitan regions of the two DESY locations, Hamburg and Zeuthen near Berlin.

\end{document}